\documentclass[preprint,aps,prb,floats]{revtex4}
\usepackage{epsfig}
\usepackage{amsmath}
\usepackage{amsfonts}
\usepackage{amssymb}
\usepackage{graphicx}

\begin{document}
\title{Energy transport and heat production in quantum engines.}
\author{ Liliana Arrachea }
\affiliation{Departamento de F{\'i}sica ``J. J. Giambiagi" FCEyN, Universidad de Buenos 
Aires, Pabell\'on I, Ciudad Universitaria (1428) Buenos Aires, Argentina.\\
lili@df.uba.ar}
\author{ Michael Moskalets }
\affiliation{ Department of Metal and Semiconductor Physics, 
NTU ``Kharkiv Polytechnic Institute'', 
61002 Kharkiv, Ukraine \\
moskalets@kpi.kharkov.ua  }
\date\today 
\begin{abstract}
A quantum dot driven by two ac gate potentials oscillating with a phase lag may be
regarded as a quantum engine, where energy is transported 
and dissipated in the form of heat. In this chapter we introduce a microscopic model
for a quantum pump and analyze the fundamental principle for the conservation of
the charge and energy in this device. We also present the basics of two well
established many-body techniques to treat quantum transport in harmonically time-dependent
systems. We discuss the different operating modes of this quantum engine, including
the mechanism of heat generation. Finally, we establish the principles of quantum
refrigeration within the weak driving regime. We also
 show that it is possible to achieve a regime
where part of the work done by some of 
the ac fields can be coherently transported and can be used
by the other driving voltages.

\end{abstract}

\maketitle

\tableofcontents

\section{Introduction}

In 1824 the french physicist Nicolas L\'eonard Sadi Carnot, better known as Sadi Carnot,  in ``Reflections on the Motive Power of Fire''
settled the principles of the modern theory of Thermodynamics by
pointing out that motive power (concept later identified as work) is due 
to the fall of caloric (concept later identified as heat) from a hot to cold body (working substance). These ideas, that provided the scientific 
support for the
technological jump based in the steam engine
 were not well understood at that time. They were actually
discovered and further elaborated thirty years later  by the German Rudolf Clausius and the British  William Thomson (Lord Kelvin).
The fundamental principles ruling the operation of the thermal engines were then summarized in the two basic laws of Thermodynamics. While the first law
simply stresses the conservation of the energy, the second one deals with the subtle distinction between a kind of energy that can be used and another
one that is dissipated in a physical process, as well as on the balance between both of them.

In 1851 Lord Kelvin also discovered Thomson effect, and showed that it was related to other thermoelectric phenomena: Peltier and Seebeck effects. Unlike 
Carnot's machines, these effects are related to non-equilibrium processes. However, they  also bring about the conceptual distinction between different 
kinds of energies:
one that is transported in some direction due to a voltage or temperature gradient, but being different from the Joule heating in the sense that the first 
one is reversible while the latter involves dissipation and, thus, irreversible effects.

\begin{figure}
\includegraphics[width=0.9\columnwidth,clip]{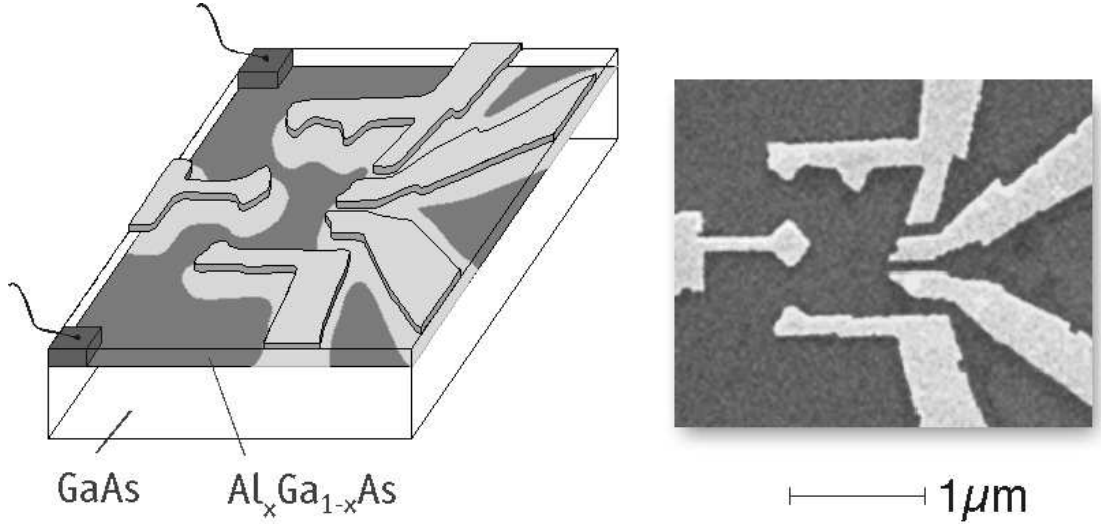}
\caption{\label{dots} Quantum dots fabricated with nanolitography in a semiconductor interface. The two leads going upwards and downwards make the contact between the dot and external reservoirs. The gate at the left can be used to apply a
voltage to  introduce a rigid shift of the  positions of the energy levels of the dot. 
The two extra leads connected at the right hand wall of the dot can be used to apply ac potentials at those points. 
Figure by Charles Marcus, Mesoscopic Lab, Harvard University.}
\end{figure}

Nowadays, we are witnessing a technological trend towards an increasing miniaturization of the electronic components. This is
 accompanied by a significant activity within communities of the
basic sciences, in the search of a better understanding of the behavior of materials and devices with sizes in the range between $1 nm$ to $10 \mu m$ as
fundamental pieces of electronic circuits. Paradigmatic examples are the quantum dots fabricated in the interfaces of semiconductor structures where 
confining gates for electrons and circuits are printed by means of nanolitography within an area of a few $\mu m^2$ (see Fig. \ref{dots}).
 Due to the small scale of these systems, they present some physical features that
 resemble the molecules. In particular, the landscape of their spectra contains 
well defined quantum levels where electrons propagate almost perfectly preserving the phase of their wave functions. 
However, they are not isolated from the external world but coupled to the substrate, gates, wires and external fields that induce the transport of electrons.
For this reason, they are classified as ``open quantum systems'' that operate out-of-equilibrium conditions. The ``external world'', instead, contains
pieces that act as macroscopic reservoirs with which the ``small quantum systems'' exchange particles and energy. Due to the mixed nature of these systems,
the concepts of classical electrodynamics and thermodynamics cannot be simply applied to them and theoretical tools that are amenable to capture their
quantum properties as well as the coupling to their environment are necessary. 

In this chapter we focus in a particular kind of devices named ``quantum pumps''. They have been realized experimentally, precisely, 
in quantum dots, where ac voltages are applied at 
their walls, and a current with a dc component is generated
in the absence of a stationary voltage difference (see Figure 1). 
From the theoretical point of view  one of the interests in
these systems is that they can be  described in terms of 
  simple Hamiltonians, including macroscopic pieces that represent the wires connected to
  the quantum dot as well as explicit terms for the 
time dependent forces that induce the transport process. The latter feature makes an important difference,
in comparison with setups where the transport is induced by means of a stationary voltage difference. This is because in such systems 
the work of the forces that keep such a bias fixed is not explicitly taken into account
in the model Hamiltonian but introduced as a boundary condition. 
 
We can imagine situations in which the 
 macroscopic wires connected at the quantum dot are at different temperatures. 
Additionally, 
 the time dependent forces do 
make work on  the system.
Thus, we can regard the quantum pump as a microscopic engine where heat can be exchanged between two sources at different temperatures,
 while work is provided
to the system and part of the energy is dissipated. 
As we will show, such an engine could even operate as a refrigerator, where there is a net
heat flow from the reservoir
at the lowest temperature to the one at the highest one. The goal of this chapter is to introduce 
theoretical tools for the 
analysis of the fundamental conservation laws at the microscopic scale, 
the explicit evaluation of the power developed by the intervening forces and the distinction 
between reversible flows and dissipated energy.   

This chapter is organized as follows. In the next section  we introduce
a simple microscopic model for a quantum pump and we will discuss the fundamental principle
of the conservation of the charge and the energy in this device. In section III
we introduce the basics of two well established many-body techniques to treat 
quantum transport in harmonically time-dependent systems. 
The first one starts from the explicit microscopic 
Hamiltonian for the system, forces and environment which is solved by recourse to non-equilibrium Green's functions. 
While the second one is based in the notion of scattering processes that the electrons experience as they cross through a quantum system under ac driving. 
Although this chapter is self-contained, we will not include a complete 
tutorial on these two techniques but we adopt a practical point of view, presenting just the main ideas while we defer the reader to more specific literature on many-body techniques for further details. 
In section IV we present explicit expressions for the energy flows in terms of the Green's functions and the scattering matrix elements introduced in section III. 
We also discus  the nature of the different components contributing to the total energy flow. 
Section V is devoted to a summary of the different operating modes that we were able to identify
in our quantum engine. Finally, in section VI we conclude with a discussion of the possible directions to extend these ideas.

\section{Background}
\subsection{Model}
\begin{figure}
\includegraphics[width=0.9\columnwidth,clip]{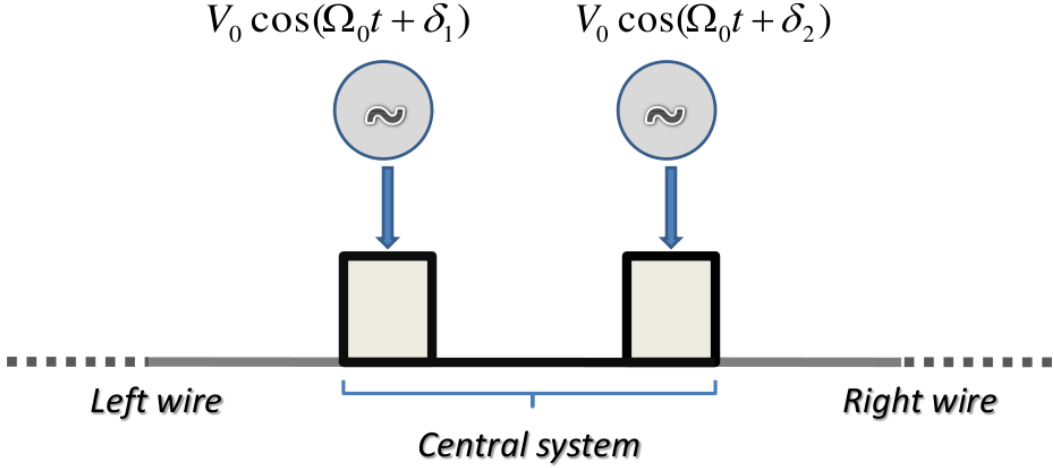}
\caption{\label{fig1} Sketch of the setup. The central system is connected two two infinite wires, which play the role of macroscopic
 reservoirs. In this example, the central system contains a profile of two barriers at which ac fields that oscillate with the same
amplitude $V_0$ and frequency $\Omega_0$, and a phase lag $\delta_2-\delta_1$. }
\end{figure}

We start by defining explicit Hamiltonians to describe the quantum electronic system as well as its environment. 
For the sake of simplicity, let us focus in a system like the one sketched in Fig \ref{fig1} where the central system $C$, on which the time-dependent forces are acting, is placed between two wires: one located at the left ($L$) and the other at the right ($R$) of $C$. 
We can identify $C$ with the quantum dot of Fig. \ref{dots}, with the two ac voltages
applied at the two extra leads connected at the wall. The $L$ and $R$ reservoirs of
Fig. \ref{fig2} correspond to the up and down ones in Fig. \ref{dots}.

Along this chapter, we make the following simplifying assumptions on the system: 
(i) We shall take into account only the two wires as the external environment for the central system and disregard other effects like, for example, the influence of the phonons of the substrate. 
Such a simplification is expected to be reasonable if we concentrate on the behavior at sufficiently low temperatures. 
(ii) We also assume that the electrons do not experience any kind of many-body interactions, like the Coulomb repulsion between electrons.
This assumption is justified in the description of the metallic wires where we can expect an efficient screening but it is justified within the system $C$ only when the structure is strongly connected to the wires and allows for the screening of the wires to penetrate into it.
In this context, the spin degrees of freedom of the electrons 
 behave independently of one another.  
For this reason, in order to simplify the notation,
 we do not consider them explicitly. In the case of considering
 them we must simply write a factor $2$ 
in front of the final expressions for the currents and densities. 
The full Hamiltonian reads:  
\begin{equation} \label{htot}
H(t)= H_L+H_C(t)+H_R+H_{c}\,,
\end{equation}
where 
\begin{equation}\label{half}
H_{\alpha}= \sum_{k \alpha } \varepsilon_{k \alpha } c^{\dagger}_{k \alpha} c_{k \alpha}\,,
\end{equation}
being $\alpha= L,R$, are Hamiltonians  of free spinless electrons that represent the wires. 
We stress that these systems are macroscopic, i.e. they contain a very large number of degrees of freedom and are in thermodynamic equilibrium. 
This means that they are completely characterized by their density of states
\begin{equation}\label{dos}
\rho_{\alpha}(\omega)= 2 \pi \sum_{k \alpha} \delta(\omega-\varepsilon_{k \alpha }/\hbar )\,,
\end{equation}
and the Fermi distribution function:
\begin{equation}\label{fermi} 
f_{\alpha}(\hbar\omega)= \frac{ 1}{e^{\beta_{\alpha}(\hbar\omega-\mu_{\alpha})}+1} \,\,,
\end{equation}
 with $\beta_{\alpha}=k_B T_{\alpha}$. 
The second term, $H_{C}(t)$, describes the quantum structure under consideration as well as the time-dependent gate potentials acting on it.
The ensuing Hamiltonian depends on the geometry of the structure as well as on the interactions that we want to take into account.
In the absence of many-body interactions, this system may be described by a single-particle Hamiltonian of the form:
\begin{equation}\label{hC1}
%{\cal H}_C({\bf r}, t) 
H_C({\bf r}, t) 
=- \frac{\hbar^2 \nabla^2}{2 m} + U({\bf r}) + \sum_{l=1}^M \delta({\bf r}- {\bf R}_l)\, e V_l(t)\,,
\end{equation}  
being $m$ the mass of an electron, which corresponds to a finite number of time-dependent potentials that we assume have the simple single-harmonic dependence: $V_l(t)= V_0 \cos(\Omega_{0} t + \delta_l)$. 
The potential $ U({\bf r})$ contains the information of the confining walls, barriers and defects of the structure. 
%As we shall see in the next section, the Hamiltonian for the central structure expressed in terms of single-particle operators is enough to formulate the scattering matrix approach to describe the flows of currents of particles and energy incoming or exiting the structure. 
%However, for 
For a detailed discussion of the conservation laws and for treating the problem with the Green's function formalism, it is convenient to express this Hamiltonian in second quantization. 
To this end we must define an appropriate single particle basis to represent the relevant operators. 
As the  structure under study occupies a reduced region of the space, it is comfortable to work with a single particle basis that is labeled by spacial coordinates, like that defined by the Wannier functions. 
It is, thus, useful to work on a discrete lattice containing a finite number ($N$) of sites and a basis of single-electron states that are localized on the lattice positions. 
The resulting Hamiltonian in second quantization corresponds to a tight-binding model. 
For simplicity, we consider this model in one-dimension (1D), although this is not an essential assumption:
\begin{equation}\label{hC2}
H_C(t)= \sum_{l=1}^N  [\varepsilon_l + eV_l (t)] c^{\dagger}_l c_l 
-w \sum_{l=1}^{N-1}  ( c^{\dagger}_l c_{l+1} + H.c)\,,
\end{equation}
where the term with $w= \langle l | (\hbar^2/2m) \partial^2/ \partial x^2 | l+1 \rangle$, being $|l \rangle $ single-electron basis state localized at the lattice position ``$l$'', describes the kinetic energy of the electrons through jumping processes between nearest-neighbor sites of the underlying lattice. 
The term with $\varepsilon_l= \langle l | U(x)   | l \rangle$, defines a static energy profile for the structure: it contains the information of the existence of barriers and wells. 
For a system with impurities, this profile can be defined in terms of a random amplitude and this model 
reduces to the Anderson model. 
The term with $V_l(t)= \langle l | \sum_{j=1}^M \delta(x_l -x_j) V_0 \cos(\Omega_{0} t + \delta_j )| l \rangle  $ represents the time dependent gates, being finite at the $M$ pumping centers and vanishing otherwise.
Finally, the term $H_c$ describes the contacts between the central system and the reservoirs. 
In our simple 1D model for the central structure the 
$L$ lead is connected to the first site, $l=1$, and the $R$ to the last site, $l=N$, of the central structure. The Hamiltonian reads:
\begin{equation}\label{hcont}
H_c=-w_{cL} \sum_{kL}(c^{\dagger}_{kL} c_1 + H.c) - w_{cR} \sum_{kR}(c^{\dagger}_{kR} c_N + H.c)\,,
\end{equation}
which describes hopping processes between the states $k \alpha $ within the wires and the points of $C$ at which the contact between the two systems is established. 

Before closing this subsection, let us mention that, depending on the physical problem under consideration, there may be other time-dependent terms in the Hamiltonian. 
For systems with ac voltages applied at the $L$ and $R$ wires, we should consider a dispersion relation with a time-dependent component in addition to the static one, $\varepsilon_{k\alpha}^0$ . 
By recourse to a gauge transformation, it can be seen that this type of ac voltages can, equivalently, be included in a time-dependent phase in the contact hopping $w_c$ (see Jauho et al. 1994). 
Another possible time-dependent term is that originated by an electric field derived from a time-dependent vector potential ${\bf A}(t)$ (see Arrachea 2002). 
That physical situation would take place, for example, when the central system is bended and closed into an annular geometry threaded by a time-dependent magnetic flux. 
In terms of the Hamiltonian this introduces a shift:
 ${\bf p} \rightarrow {\bf p} - (e/c) {\bf A}(t)$ in the momentum of the electrons, being $c$ the velocity of light. 
In the tight-binding basis this translates into time-dependent phases in the hopping parameter $w$ of $H_C(t)$ along with the periodic boundary condition $N+1 \equiv 1$. 
Finally, the study of the coupling to external classical radiation fields is usually treated   within the so called dipolar approximation, which in our second quantization language results in a diagonal voltage profile $V_j(t)$ as in  Eq.\,(\ref{hC2}) with $V_j \sim j V_0$, with  $V_0$ constant and $\delta_j=\delta, \forall j$ (see Kohler et al. 2005).

\subsection{Conservation laws and instantaneous currents} \label{conslaw}
\subsubsection{Particle currents and the conservation of the charge}
A consistent way to define expressions for the electronic currents along the different pieces of the structure is starting from the evolution of the electronic density 
 $n_l = c^{\dagger}_l c_l$. The variation of $n_l$ is due to the difference between the charge flow exiting and entering the infinitesimal volume that encloses that point:
\begin{equation}
-e \frac{d}{dt} \langle n_l \rangle = J_{l}(t) - J_{l-1}(t)\,,
\end{equation}
where $J_{l}(t)$ denotes the current exiting the site $l$ towards the neighboring site $l+1$. 
We denote with a positive sign the flows pointing from left to right. 
The variation in time of the local charge  can be calculated within the Heisenberg picture by recourse to the Eherenfest theorem:
\begin{equation}
e\frac{d}{dt} \langle n_l \rangle = - \frac{i e}{\hbar}  \langle [H, c^{\dagger}_l(t) c_l(t) ] \rangle\,.
\end{equation}
Thus, the explicit evaluation of the above commutator defines an explicit expression for the current.
If we consider a  site within $C$ we obtain:
\begin{equation}\label{jl}
J_l(t)= \frac{i e w }{\hbar} \langle   c^{\dagger}_l(t) c_{l+1}(t)- c^{\dagger}_{l+1}(t) c_{l}(t) \rangle\,.
\end{equation}
It is easy to verify that the above expression coincides with the mean value of the operator $e {\bf v}$, being ${\bf v}$ the velocity expressed in second quantization in the basis of localized functions.  
Similarly, if we consider the contact between $C$ and one of the reservoirs we get the following expression for the current that exits the reservoir $\alpha$:
\begin{equation} \label{jal}
J_{\alpha}(t)= \frac{i e w_{c\alpha} }{\hbar} \sum_{k \alpha} \langle  
c^{\dagger}_{k \alpha}(t) c_{l \alpha}(t) - c^{\dagger}_{l\alpha}(t) c_{k \alpha }(t) \rangle\,,
\end{equation}
where $l \alpha = 1, N $ for $\alpha=L, R$. 

\subsubsection{Energy currents, power, and the conservation of the energy}

\begin{figure}
\includegraphics[width=0.9\columnwidth,clip]{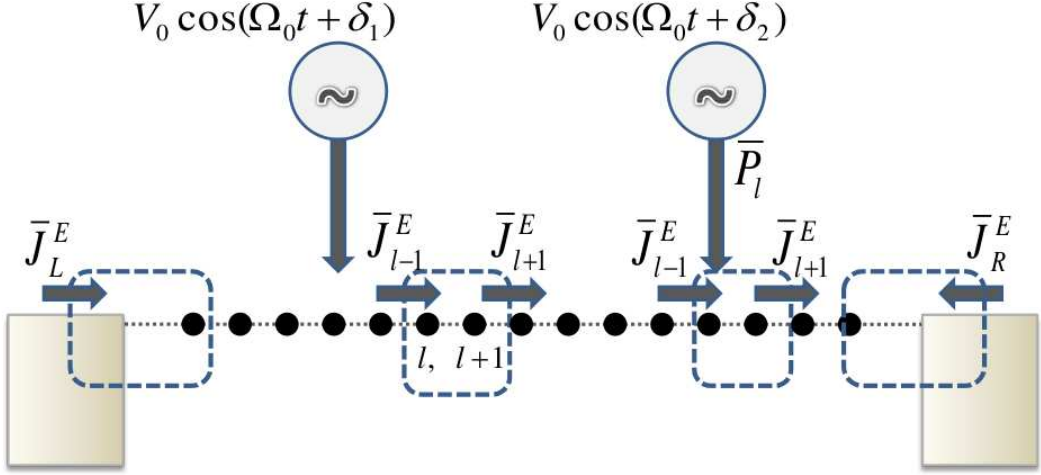}
\caption{\label{fig2} Analysis of  the energy balance in our setup.
The dashed boxes
enclose the elementary volume of the system where the evolution of the energy is studied. It
contains two sites of the underlying lattice. The arrows  indicate the direction
 that we have defined as positive for energy currents  along the different pieces of the system
as well as the powers done by the fields.}
\end{figure}
In order to define energy currents, we proceed along a similar line as before. 
In this case, we analyze the evolution of the energy density at a given elementary volume and write the equation of the conservation of the energy. 
As our Hamiltonian $H_C(t)$ contains terms involving  positions up to nearest-neighbors, the smallest volume for analyzing the evolution of the energy density in our 1D lattice is that enclosed by a box confining a bond of nearest-neighbor sites (see Figure \ref{fig2}). 
If we denote by $E_{l,l+1}$ the total energy stored within such a box, the equation for the conservation of the energy reads:
\begin{equation}
\frac{d E_{l,l+1}}{dt}=  J^E_{l+1}(t) - J^E_{l}(t) + P_l(t) + P_{l+1}(t)\,,
\end{equation}
where the first two terms denotes the difference between outgoing (from $l+1$ to $l+2$) and incoming (from $l-1$ to $l$) flows, with the same sign convention as in the case of charge flows, while the last two terms denote the power done by the external fields, which are defined as positive when it is provided by the forces. 
The latter terms vanish if the time-dependent gate potentials are not acting at the points $l$ and $l+1$ enclosed by the box. 
Our box can also enclose the contact bond between the reservoir $\alpha$ and the central system (see Fig. \ref{fig2}), in which case we get:
\begin{eqnarray}
\frac{d E_{L,1}}{dt}&=&  J^E_{1}(t) - J^E_L(t) + P_1(t)\,,\nonumber \\
\frac{d E_{N,R}}{dt}&=& -J^E_R(t) -  J^E_{N-1}(t) + P_N(t)\,,
\end{eqnarray}
where $J^E_{\alpha}(t)$ denotes the energy flow that exits the reservoir $\alpha$. 
As in the previous subsection, the explicit expressions for the flows and powers are derived by recourse to the Eherenfest theorem:
\begin{eqnarray}
\frac{d E_{l,l+1}}{dt} & = &  - \frac{i }{\hbar}  \langle \big[H, ( \varepsilon_l(t) c^{\dagger}_l(t) c_l(t) +
\varepsilon_{l+1}(t) c^{\dagger}_{l+1}(t) c_{l+1}(t) -w\{ c^{\dagger}_l(t) c_{l+1}(t)+c^{\dagger}_{l+1}(t) c_{l}(t) \} ) \big] \rangle \nonumber \\
& & +  e \frac{d V_l(t)  }{d t} \langle  c^{\dagger}_l(t) c_l(t) \rangle + 
e \frac{d V_{l+1}(t) }{d t}  \langle  c^{\dagger}_{l+1}(t) c_{l+1}(t) \rangle\,,
\end{eqnarray}
with $\varepsilon_l(t)=\varepsilon_l^0 + eV_l(t)$, and a similar expression for the volume enclosing the contact bonds between $C$ and the reservoirs.
From the evaluation of the previous terms, we obtain the explicit expressions for the energy currents:
\begin{eqnarray}
J^E_{l+1}(t) & = &   \frac{i w }{\hbar}[w \langle c^{\dagger}_l(t) c_{l+2}(t)- c^{\dagger}_{l+2}(t) c_{l}(t) \rangle -  
\varepsilon_{l+1}(t) \langle c^{\dagger}_{l+1}(t) c_{l+2}(t)- c^{\dagger}_{l+2}(t) c_{l+1}(t) \rangle]\,,  \label{jel} \\
 J^E_{\alpha}(t) & =& \frac{i w_{c\alpha} }{\hbar} \sum_{k \alpha}\varepsilon_{k \alpha}  \langle c^{\dagger}_{k \alpha}(t) c_{l \alpha}(t)
- c^{\dagger}_{l \alpha}(t) c_{k \alpha }(t) \rangle\,, \label{jeal}
\end{eqnarray}
and power developed by the ac voltages:
\begin{equation} \label{power}
P_l(t)= e \frac{d V_l(t)  }{d t} \langle  c^{\dagger}_l(t) c_l(t) \rangle\,.
\end{equation}

\subsection{Continuity equations, dc charge and energy currents and mean powers}\label{curdef}
In the absence of sinks and sources, the average of the charge enclosed by any volume of the sample over one cycle with of period $\tau= 2 \pi /\Omega_{0}$ must remain constant, which defines the following continuity equation for the dc charge current (microscopic Kirchoff law):
\begin{equation}\label{contch}
\overline{J}_{l}=\overline{J}_{l'}= J\,,
\end{equation}
for arbitrary $l, l'$ along $C$, where we denote $\overline{A}= 1/\tau \int_0^{\tau} dt A(t)$.

Analogously, for any volume enclosing lattice points running from 
$l+1, \ldots, l'$ we get:
\begin{equation} \label{conten}
\overline{J}^E_{l} = \overline{J}^E_{l'} + \sum_{j=l+1}^{l'} \overline{P}_j\,,
\end{equation}
where the last term defines the power done by all the voltages enclosed by the volume. The above equation
 reduces to a ``Kirchoff law'' for the dc energy current when we enclose a region that is free from the time-dependent voltages, 
in which case the last term vanishes.

\subsection{Heat current} \label{heat}
At this point, it is important to mention that the dc energy current $\overline{J}^E$ defined above does not necessarily coincide with the heat current which we will denote $J^Q$. 
This is because what is understood as ``heat'' is usually the energy transferred from one system to another as a consequence of a temperature difference. 
In order to understand this difference, let us consider that our reservoirs are at temperature $T=0$, let us place our volume enclosing the contact between one reservoir and the central system and let us assume that we are not enclosing an ac local voltage within it. 
Now, let us analyze the following picture based on heuristic  arguments that we shall better formalize it in subsequent sections. 
Let us assume that the ac voltages are so weak in amplitude and oscillate with such a low frequency that we can disregard the power they develop. 
Let us assume that, anyway, they are able to move a small portion of electrons with energies very close to the Fermi energy of the reservoirs $\mu$, that we recall is the same for $L$ and $R$. 
This weak motion give place to currents of particles and energy which, from  the definitions (\ref{jal}) and (\ref{jeal}), are approximately related through $J^E_{\alpha}(t) \sim (\mu/e) J_{\alpha}(t)$. 
The same relation holds for $L$ as well as $R$ reservoirs and a small current of particles may translate in a large current of energy, since $\mu$ can be large. 
The dc component of the charge current may be finite and should be positive in one reservoir and negative in the other one, indicating that there is a net flow of charge between $L$ and $R$ or vice versa. 
The above relation, therefore, tells us that there is a concomitant net flow of energy from a reservoir to the other one, in spite of the fact that we are assuming that both reservoirs are at zero temperature. One could complain  that we have not taken into account the power done by the time-dependent fields 
which would tend to heat the system. 
However, let us recall that an appropriate choice of $\mu$ would allow us an arbitrary large value of the energy flow, against which we can disregard a contribution like (\ref{power}) for weak and slow ac voltages. 
In summary, the above considerations lead us to conclude that there may exist a net energy flow which cannot be identified as ``heat'' but has rather a convective nature. 
In order to better quantify heat, it is thus natural to subtract from the energy flow, the convective component $(\mu/e) J$, i.e.:
\begin{equation}\label{heatc}
J^Q_{l}= \overline{J}^E_{l}- \frac{\mu}{e} J\,, \;\;\;\;\;\;\;\;    J^Q_{\alpha}= \overline{J}^E_{\alpha}-\frac{\mu}{e}J\,.
\end{equation}
Notice that, while the convective term is constant along all the pieces of the system due to the continuity of the charge, the heat and energy currents may have different values due to the contribution of the power done by the external voltages, see Eq.\,(\ref{conten}).

To give a more formal definition of the heat flow $J^{Q}_{\alpha}$, let us consider the particle
 and the energy balances for a given reservoir $\alpha$ that
 is kept at fixed both the chemical potential $\mu_{\alpha}$ and the temperature $T_{\alpha}$. 
If the charge current $\overline{J}_{\alpha}$ and the energy flow $\overline{J}^{E}_{\alpha}$
 enter the reservoir $\alpha$, then the number of particles and the energy of a reservoir 
should change. 
This, in turn, would change the chemical potential and the temperature of a reservoir.
 In order to 
 keep $\mu_{\alpha}$ fixed, the electrons have to be removed with the rate $\overline{J}_{\alpha}/e$ out of the reservoir. 
Therefore, while keeping $\mu_{\alpha}$ constant we necessarily remove energy with rate $\mu_{\alpha} \overline{J}_{ \alpha}/e$. 
We stress that the convective energy $\mu_{\alpha} \overline{J}_{ \alpha}/e$ is taken out at equilibrium conditions, therefore, it can be reversibly given back. 
In general this energy does not coincide with energy flow $\overline{J}^{E}_{\alpha}$ entering the reservoir. 
To prevent heating of reservoir one needs additionally to take out the energy with rate $J^{Q}_{\alpha} = \overline{J}^{E}_{\alpha} - \mu_{\alpha} \overline{J}_{ \alpha}/e$ without 
taking out particles.
Since the reservoir can not produce work, the only  way to remove the remaining energy is 
to put it in contact with other large body playing the role of a thermostat. 
The energy exchange between the reservoir $\alpha$ and the thermostat is 
essentially irreversible. For this reason we interpret this part of the total energy 
as ``heat'' and  we identify $J^{Q}_{\alpha}$ as the heat flow.   
If the thermostat would be absent the temperature of a reservoir would change. 
As we will show $J^{Q}_{\alpha}$ can be directed either to the reservoir from the central system or back. 
Hence, in the absence of a thermostat, the reservoir can be either heated or cooled. 
We stress that the energy transported at the rate $J^{Q}_{\alpha}$ becomes ``heat''
 only deep inside the macroscopic reservoir, where the electrons 
scattered by the dynamical central system, are able to equilibrate.

\section{State of the art}

In this section we briefly review the many-body techniques to evaluate the currents and powers defined in the previous section.

\subsection{Green's functions formalism}

\subsubsection{Expectation values of observables and Green's functions}\label{exp}

The expectation value of any one-body observable, 
 $\langle A(t) \rangle  = \sum_{l,l'} \langle l' |A(t) |l \rangle  \langle c^{\dagger}_{l'}(t) c_l(t) \rangle $, can be regarded as follows:
 \begin{equation}
\langle \hat A(t) \rangle  
= -i \lim_{t'\rightarrow t} \sum_{l,l'} \langle l' |A(t) |l \rangle\, G^{<}_{l,l'}(t,t')\,,
\end{equation}
being 
\begin{equation}\label{lesser}
G^<_{l,l'}(t,t')= i\, \langle c^{\dagger}_{l'}(t') c_{l}(t) \rangle\,,
\end{equation}
a ``lesser'' Green's function. Our goal, now, is to derive equations for the evolution of 
this Green's function and strategies to solve them.
 
\subsubsection{Brief review of the theory of the non-equilibrium Green's functions}
The formal theory of non-equilibrium Green's function has been developed independently by Kadanoff and Baym [Kadanoff and Baym 1959], Schwinger [Schwinger 1961] and Keldysh [Keldysh 1962]. 
The structure of that theory is very similar to the one of causal Green's functions at zero temperature (see Mahan 1990), except for the fact that in non-equilibrium situations, the assumption that the state of the system at time $+\infty$ differs just in a phase from the state in $-\infty$ does not longer holds. 
The way to overcome this inconvenience is to define the evolution along a special contour ${\cal C}$ that defines a round trip, first going from $-\infty$ to $+\infty$ and then going back to $-\infty$. 
As in equilibrium problems, the precise description of that evolution can be accomplished with the help of Wick's theorem and Feynman diagrams and one of the big powers of this technique is the possibility of treating
many-body interactions in a systematic way.  

We skip here the technical just highlighting the main ideas leading to some useful identities, and we defer the reader to more specialized literature (see Mahan 1990, Haug and Jauho 1996, Caroli et al. 1971, Pastawski 1992, Jauho et al. 1994).
Instead of the time-ordering operator used in the equilibrium theory, it is convenient to work with
contour-ordered Green's functions:
\begin{equation}
G(1,1')= -i\, \langle T_{\cal C}[c(1)c^{\dagger}(1')] \rangle\,,
\end{equation}
where $1,1'$ is a schematic notation that labels the electronic degrees of freedom and
time in the same index and the operator  $T_{\cal C}$ denotes time-ordering along a contour
that begins in $-\infty$ evolves to $+\infty$ (${\cal C}_+$) and then turns  back 
from $+\infty$  to  $-\infty$ (${\cal C}_-$). This function corresponds to the 
casual ``time-ordered'', the ``lesser'', the ``anti-time-ordered'' or ``greater'' function depending
on the position of the two times along the closed time-contour:
\begin{eqnarray}
& & G(1,1') =  G_c(1,1')\,, \; t_1,t_{1'} \in {\cal C}_+\,, \;\;\;\;\;\;\;\;\;\;\; 
G(1,1') =  G^<(1,1')\,, \; t_1,\in {\cal C}_+\,, t_{1'} \in {\cal C}_-\,, \nonumber \\
& & G(1,1') =  G_{\overline c}(1,1')\,, \; t_1,t_{1'} \in {\cal C}_-\,, \;\;\;\;\;\;\;\;\;\;\; 
G(1,1') =  G^>(1,1')\,, \; t_1,\in {\cal C}_-\,, t_{1'} \in {\cal C}_+\,,
\end{eqnarray}
with: 
\begin{eqnarray} \label{greenc}
& &G_c(1,1')= -i\,\Theta(t_1-t_{1'}) \langle c_1(t_1) c^{\dagger}_1(t_1') \rangle 
 +i\,\Theta(t_{1'}-t_1 ) \langle c^{\dagger}_1(t_1')  c_1(t_1) \rangle\,, \nonumber \\
& &G_{\overline c}(1,1')= -i\,\Theta(t_{1'}-t_1 )\langle c_1(t_1) c^{\dagger}_1(t_1') \rangle 
 +i\,\Theta(t_1-t_{1'})  \langle c^{\dagger}_1(t_1')  c_1(t_1) \rangle\,, \nonumber \\
& &G^<(1,1')=i\, \langle c^{\dagger}_1(t_1')  c_1(t_1) \rangle\,,
\;\;\;\;\;\;\;\;\;\;\;\;\;\;\; G^>(1,1')=-i \,\langle c_1(t_1) c^{\dagger}_1(t_1')   \rangle\,,
\end{eqnarray}
which are not independent functions, but satisfy: $G_c+G_{\overline c}=G^<+G^>$. It is also 
convenient
to define ``retarded'' and ``advanced'' functions:
\begin{eqnarray}\label{gra}
G^R(1,1')& =& \Theta(t_1-t_{1'})[G^>(1,1')-G^<(1,1')]\,,\nonumber \\
G^A(1,1')&=& \Theta(t_{1'}-t_1)[G^<(1,1')-G^>(1,1')]\,,
\end{eqnarray}
which are also related through $G^R-G^A=G^>-G^<$, $G^A(1,1')= [G^R(1',1)]^*$ and
$G^{<}(1,1')= - [G^{>}(1',1)]^*$.

In our simple model of non-interacting electrons, we can split by convenience the Hamiltonian
in two parts $H(t)=H_0(t)+H'(t)$, with both terms being of one-body type but $H_0$ being
easily solved.
 The evolution of this Green's function is given by Dyson's equation, which for
a one-body Hamiltonian reads:
\begin{equation} \label{dym}
G(1,1')=G_0(1,1')+\hbar^{-1}\int_{\cal C} d^3x_2 dt_2 G(1,2) H'(2) G_0(2,1')\,.
\end{equation}
Notice that the above equation actually represents a matricial integral equation
if we distinguish the positions of the times as in (\ref{greenc}).
A convenient tool to derive explicit equations for the different components (\ref{greenc})
is a theorem due to Langreth, which states given a product of contour-ordered Green's functions
of the form: 
\begin{equation}
G(t_1,t_{1'})= \int_{\cal C} dt_2 G_1(t_1,t_2)G_2(t_2,t_{1'})\,,
\end{equation}
then, the following relations hold for the different components:
\begin{eqnarray}\label{lang}
G^{R,A}(t_1,t_{1'}) & = &  \int_{t_{1'}}^{t_1} dt_2 
G_1^{R,A}(t_1,t_2)G_2^{R,A}(t_2,t_{1'})\,, \nonumber \\
G^{<,>(}t_1,t_{1'}) & = &  \int_{-{\infty}}^{-{\infty}} dt_2 [G_1^R(t_1,t_2)G_2^{<,>}(t_2,t_{1'})+G_1^{<,>}(t_1,t_2)G_2^A(t_2,t_{1'})]\,.
\end{eqnarray}
Therefore, if we want to compute $G^<$, in order to evaluate expectation values of observables, we must also solve
two coupled equations for that function and $G^R$.

\subsubsection{Green's functions and Dyson's equations in our problem}
Let us first split our Hamiltonian as follows: $H_0(t)=H_L+H_R+H_C(t)$ and $H'=H_{c}$.
The Dyson's  equation for the retarded function reads:
\begin{equation} \label {dyint1}
G_{j,j'}(t,t')=g^{0}_{j,j'}(t,t')+ \hbar^{-1}\sum_{j_1}\int_{\cal C} dt_1 G_{j,j_1}(t,t_1) 
H_{c}  
g^{0}_{j_1,j'}(t_1,t')\,,
\end{equation} 
where $g^0_{j,j'}(t,t')$ is  the contour-ordered Green's function of $H_0(t)$ and $j,j'$ run over 
all the electronic 
degrees of freedom of this Hamiltonian.
If we write the above equation explicitly for one of indexes in the reservoir  
and $l \in C$:
\begin{equation} \label{dyint2}
G_{l,k \alpha}(t,t')=-w_c\int_{\cal C} dt_1 G_{l,l\alpha}(t,t_1)
g^{0}_{k \alpha,k \alpha }(t_1,t')\,,
\end{equation}
For the two indexes $l,l' \in C$:
\begin{eqnarray} \label{dyint3}
G_{l,l'}(t,t') & = & g^{0}_{l,l'}(t,t')
-w_c \sum_{\alpha=L,R}  \int_{\cal C} dt_1 G_{l,k \alpha}(t,t_1) g^{0}_{l \alpha,l'}(t_1,t')\,,
\nonumber \\
 & = & g^{0}_{l,l'}(t,t')+ \hbar^{-1}\sum_{\alpha=L,R}
 \int_{-\infty}^{\infty} dt_1 dt_2 G_{l,l\alpha}(t,t_1) 
\Sigma_{\alpha}(t_1,t_2)  g^{0}_{l \alpha, l'}(t_2,t')\,,
\end{eqnarray}
where going from the first to the second identity we have substituted (\ref{dyint2}). We have also
defined the ``self-energy'':
\begin{equation}\label{sigma}
\Sigma_{\alpha}(t_1,t_2)=|w_{c \alpha}|^2 \sum_{k \alpha} g^{0}_{k \alpha,k \alpha }(t_1,t_2)\,.
\end{equation}

In order to evaluate the currents that we have defined in section (\ref{curdef}), we need $G^<_{l,l'}(t,t')$ for $l,l' \in C$ 
and $G^{<,>}_{k \alpha, l\alpha}(t,t')$. Applying the Langreth's rules (\ref{lang}) to the above equations we get
(see Arrachea 2002 and 2005):
\begin{eqnarray} \label{glesc}
G^{<,>}_{l,k \alpha}(t,t')& = & -w_{c \alpha}\int_{-\infty}^{+\infty} dt_1 
[G^R_{l,l\alpha}(t,t_1) g^{0,<,>}_{k \alpha,k \alpha }(t_1,t') +
G^{<,>}_{l l\alpha }(t,t_1)g^{0,A}_{k \alpha,k \alpha }(t_1,t') ]\,,
%\nonumber \\
%G^{>}_{l,k \alpha}(t,t')& = & -w_{c \alpha}\int_{-\infty}^{+\infty} dt_1 
%[G^R_{l,l\alpha}(t,t_1) g^{0,>}_{k \alpha,k \alpha }(t_1,t') +
%G^>_{l l\alpha }(t,t_1)g^{0,A}_{k \alpha,k \alpha }(t_1,t') ],
\end{eqnarray}
along the contact and 
\begin{eqnarray}\label{glesin}
G^{<,>}_{l, l'}(t,t')& = & \hbar^{-1}\sum_{\alpha} \int_{-\infty}^{+\infty} dt_1 dt_2
G^{R}_{l,l \alpha }(t,t_1) \Sigma^{<,>}_{\alpha}(t_1,t_2) G^{A}_{l \alpha, l' }(t_2,t')\,, \nonumber \\
%G^>_{l, l'}(t,t')& = & \sum_{\alpha} \int_{-\infty}^{+\infty} dt_1 dt_2
%G^{R}_{l,l \alpha }(t,t_1) \Sigma^>_{\alpha}(t_1,t_2) G^{A}_{l \alpha, l' }(t_2,t'),
\end{eqnarray}
for coordinates $l,l'$  belonging to the central system $C$. The latter equation is obtained after some algebra
and after dropping a term that contains $g_{l,l'}^{0,<}(t,t')$ which can shown to relevant only in the description of transient behavior (see
Jauho et al. 1994).
The different components of $G^{0}(t,t')$ within the reservoirs are straightforwardly evaluated:
\begin{eqnarray}\label{gres} 
g^{0,R }_{k \alpha,k \alpha}(t_1,t_2) & = & -i\, \Theta(t_1-t_2) \exp\{-i (\varepsilon_{k \alpha}/\hbar)(t_1-t_2) \}\,, \nonumber \\
g^{0,<,>}_{k \alpha,k \alpha}(t_1,t_2) & = & \lambda_{\alpha}^{<,>}(\varepsilon_{k \alpha})
% i   f_{\alpha}(\varepsilon_{k \alpha}) 
\exp\{-i (\varepsilon_{k \alpha}/\hbar) (t_1-t_2)\}\,,
%\nonumber \\
%g^{0,>}_{k \alpha,k \alpha}(t_1,t_2) & = &  -i 
%[1- f_{\alpha}(\varepsilon_{k\alpha})]  \exp\{-i \varepsilon_{k \alpha} (t_1-t_2)\},
\end{eqnarray}
with $\lambda_{\alpha}^{<}(\varepsilon_{k \alpha})=i   f_{\alpha}(\varepsilon_{k \alpha})$ and 
$\lambda_{\alpha}^{>}(\varepsilon_{k \alpha})= -i\, [1- f_{\alpha}(\varepsilon_{k \alpha})]$.
With these functions, it is possible to obtain expressions for the different components of
(\ref{sigma}) in terms of the density of states of the reservoir $\rho_{\alpha}(\omega)$ 
given in (\ref{dos}):
\begin{eqnarray}\label{sigmas} 
\Sigma_{\alpha}^R(t_1,t_2) & = & -i \Theta(t_1-t_2) \int_{-\infty}^{+\infty}
\frac{ d \omega}{2 \pi} e^{-i \omega (t_1-t_2)} \Gamma_{\alpha}(\omega)\,, \nonumber \\
\Sigma_{\alpha}^{<,>}(t_1,t_2) & = &   \int_{-\infty}^{+\infty}
\frac{ d \omega}{2 \pi} e^{-i \omega (t_1-t_2)}\lambda^{<,>}_{\alpha}(\hbar\omega)
 \Gamma_{\alpha}(\omega)\,, 
%\nonumber \\
%\Sigma_{\alpha}^>(t_1,t_2) & = & -i  \int_{-\infty}^{+\infty} 
%\frac{ d \omega}{2 \pi} e^{-i \omega (t_1-t_2)}[ 1- f_{\alpha}(\omega)] \Gamma_{\alpha}(\omega),
\end{eqnarray}
with $\Gamma_{\alpha}(\omega)=|w_{c\alpha}|^2 \rho_{\alpha}(\omega)$.

To evaluate (\ref{glesc}) and (\ref{glesin}) we still have to calculate the retarded function within the system $C$ (recall that the
advanced function can be obtained from $G^A_{l,l'}(t,t')= [G^R_{l',l}(t',t)]^*$). The equation for the
retarded function is the retarded component of (\ref{dyint3}) and can be derived by applying Langreth
rules on this equation. The result is:
\begin{equation} \label{gretc}
G^R_{l, l'}(t,t') =  g^{0,R}_{l, l'}(t,t')+  \hbar^{-1}\sum_{\alpha} \int_{t'}^{t} dt_1 dt_2
G^{R}_{l,l \alpha }(t,t_1) \Sigma^R_{\alpha}(t_1,t_2) g^{0,R}_{l \alpha, l' }(t_2,t')\,,
\end{equation}
with $ g^{0,R}_{l, l'}(t,t')$ being the retarded Green's function of the system described only by
 $H_C(t)$ isolated from the reservoirs. This function is, in turn still an unknown in our problem.
Instead of explicitly evaluating it, we find it more convenient to operate with (\ref{gretc})
in order to find an equivalent equation for $\hat{G}^R(t,t')$ as follows.
We first derive  an equation of motion for $ g^{0,R}_{l, l'}(t,t')$,
starting from
the very definition of the retarded function, see Eq.\,(\ref{gra}),
and by writing the evolution of $c_j(t)$ 
with $H_C(t)$ in the Heisenberg representation:
%(we now put $\hbar=1$ and reintroduce it in the final expressions for the current):
\begin{equation}
i \hbar \stackrel{\cdot}{c}_j(t)= [H_C(t),c_j(t)]\,,
\end{equation}
we get:
\begin{equation}
 - i \hbar \frac{\partial}{\partial t'} \hat{g}^{0,R}(t,t') - \hat{g}^{0,R}(t,t') \hat{H}_C(t')   =  \hbar \hat{1} \delta (t-t')\,,
\end{equation}
where $\hat{g}^{0,R}(t,t')$ is a $N \times N$ matrix with elements $g^{0,R}_{l, l'}(t,t')$ and $ \hat{1} $ is the
 $N \times N$
 identity matrix.
The above equation means that  $\{-i \hbar \partial/\partial t'- \hat{H}_C(t') \}=
[\hat{g}^{0,R}]^{-1}$. Therefore, we have not evaluated explicitly the function
$\hat{g}^{0,R}(t,t')$ but we have identified an operator which is its inverse.
We act 
 with this operator from the right of (\ref{gretc}) and we consider the following  
splitting of the central Hamiltonian $H_C(t)= H^0_{C} + H'_{C}(t)$, where $ H_{C}(t)$ collects all the explicit time-dependent terms of $H_C(t)$ and
$H^0_{C}$, the remaining ones. We get:
\begin{equation}
 -i\hbar \frac{\partial}{\partial t'}  \hat{G}^{R}(t,t') - \hat{G}^{R}(t,t') \hat{H}^0_{C}  - \int dt_1 \hat{G}^{R}(t,t_1) \hat{\Sigma}^R(t_1,t') - \hat{G}^{R}(t,t') H'_{C}(t')=  \hbar \hat{1} \delta (t-t')\,,
\end{equation}
being $\Sigma^R_{l,l'}(t,t')=\sum_{\alpha} \delta_{l,l\alpha} \delta_{l',l\alpha} \Sigma_{\alpha}^R(t,t')$, the matrix elements of $\hat{\Sigma}(t,t')$.
We define a function $\hat{G}^{0,R}(t,t')$, such that 
 $[\hat{G}^{0,R}]^{-1}=\{-i \hbar \partial/\partial t'- \hat{H}^0_C - \hat{\Sigma}^R\}$:
\begin{equation}\label{g0}
 -i \hbar \frac{\partial}{\partial t'}  \hat{G}^{0,R}(t,t') - \hat{G}^{0,R}(t,t') \hat{H}^0_{C}  - 
\int dt_1 \hat{G}^{0,R}(t,t_1) 
\hat{\Sigma}^R(t_1,t')  = \hbar \hat{1} \delta (t-t')\,.
\end{equation}
Multiplying (\ref{gretc}) by the right with $\hat{G}^{0,R}$, we finally find
the following equation
for the full retarded Green's function 
\begin{equation}\label{gretf1}
\hat{G}^R (t,t') =  \hat{G}^{0,R}(t,t')+  \hbar^{-1}\int_{t'}^{t} dt_1 
\hat{G}^{R}(t,t_1) H'_{C}(t_1) \hat{G}^{0,R}(t_1,t')\,.
\end{equation}
This equation is completely equivalent to (\ref{gretc}) but has the advantage that
the function $\hat{G}^{0,R}$ is an equilibrium Green's function, which evolves according
to the stationary terms of the full Hamiltonian $H$. 
The above expression has a structure
which is particularly adequate for perturbative solutions in the time dependent part of
$H_C(t)$. We shall exploit this property later.

Equation (\ref{g0}) can be easily solved by performing the Fourier transform:
$\hat{G}^{0,R}(\omega)= \hbar^{-1}\int_{0}^{\tau} d \tau e^{-i (\omega + i \eta) \tau} \hat{G}^{0,R}(\tau)$, with $\eta >0$,
 since, as we have mentioned before, it corresponds to an equilibrium Green's function that depends on $t-t'$.
The result is : 
\begin{equation}
\hat{G}^{0,R}(\omega)=[\hbar(\omega + i\eta) \hat{1}- \hat{H}^0_{C} - \hat{\Sigma}^R(\omega)]^{-1}\,,
\end{equation}
and can be explicitly evaluated by simply inverting the above $N\times N$ complex matrix.
 Substituting in (\ref{gretf1}) and
performing a Fourier transform in $t-t'$, Eq.\,(\ref{gretf1}) results for our specific Hamiltonian (\ref{hC2}):
\begin{eqnarray}\label{gretw}
\hat{G}^R (t,\omega) & = &  \hat{G}^{0,R}(\omega)+ \sum_{k=\pm}e^{-i k \Omega_{0} t} 
\hat{G}^R (t,\omega+ k \Omega_{0} ) 
\hat{V}(k) \hat{G}^{0,R}(\omega)\,,
% \nonumber \\
%& & + e^{i\Omega_{0} t} \hat{G}^R (t,\omega-\Omega_{0} ) \hat{V}(-1) \hat{G}^{0,R}(\omega),
\end{eqnarray}
where the matrix $\hat{V}(1)$ contains elements $V_{l,l'}=\delta_{l,l'} \sum_{j=1}^M \delta(x_l-x_j) eV_0 e^{-i\delta_j}$ and $\hat{V}(-1)=[\hat{V}(1)]^*$.
The  linear set (\ref{gretw}) has the same structure as the dynamics of a problem in which electrons 
with a given energy $\hbar\omega$
interact with a potential $V$ emitting or absorbing an energy quantum $\hbar\Omega_{0}$ and 
scatter with a final energy
$\hbar\omega \pm \hbar\Omega_{0}$.  
The solution of (\ref{gretw}) leads to the complete solution of the problem.

Due to the harmonic dependence on the time $t$ of these equations, the retarded Green's function 
can 
be expanded in a Fourier series
as follows: 
\begin{equation}\label{gfou}
\hat{G}^R (t,\omega)=\sum_{n=-\infty}^{+\infty} \hat{\cal G}(n,\omega) e^{-i n \Omega_{0} t}\,.
\end{equation}   
We give the name of {\em Floquet component } to the functions $ \hat{\cal G}(n,\omega)$,
because  (\ref{gfou}) has a similar structure as that proposed by
 Floquet  for the  wave functions of time-periodic
Hamiltonians. The different components obey the following useful identity:
\begin{equation}\label{prop}
\hat{\cal G}(n,\omega)-\hat{\cal G}^{\dagger}(-n,\omega_n )= -i \sum_{n'}
\hat{\cal G}(n+n',\omega_{-n'})\hat{\Gamma}(\omega_{-n'}) 
\hat{\cal G}(n',\omega_{-n'})^{\dagger}\,,
\end{equation}
where we have introduced the following notation $\omega_n=\omega+n \Omega_{0}$. To prove (\ref{prop})
we start from the definition (\ref{gra}) of the retarded Green's function
for indexes $l,l'\in C$.  Replacing (\ref{glesin}) and inserting there 
the representation (\ref{gfou}), we get:
\begin{equation}
\hat{G}^R(t,t')=-i\hbar \Theta(t-t') \sum_{k_1,k_2}
\int_{-\infty}^{+\infty} \frac{ d\omega}{2 \pi} e^{-i [\omega (t-t') + 
\Omega_{0}(k_1 t -k_2 t')]}
\hat{\cal G}(k_1,\omega)\hat{\Gamma}(\omega) \hat{\cal G}^{\dagger}(k_2,\omega)\,,
\end{equation} 
where $\hat{\Gamma}(\omega)$ contains as matrix element ${\Gamma}_{l,l'}(\omega)
= \delta_{l,l \alpha} \delta_{l',l \alpha} \Gamma_{\alpha}(\omega)$. Calculating the 
 Fourier transform of this function with respect to $t-t'$ and collecting the $n$-th Fourier
coefficient (\ref{gfou}) we find:
\begin{equation}
\hat{\cal G}(n,\omega)= \sum_{n'}
\int_{-\infty}^{+\infty} \frac{ d\omega'}{2 \pi} \frac{
\hat{\cal G}(n+n',\omega')\hat{\Gamma}(\omega') \hat{\cal G}(n',\omega')^{\dagger}}
{\omega - \omega'_{n'}+ i\eta}\,, \;\;\;\;\;\;\;\; \eta >0\,,
\end{equation}
which leads to the identity (\ref{prop}) using:
\begin{equation}
\frac{1}{\omega - \omega' + i \eta}= {\cal P}\left\{ \frac{1}{\omega -\omega' } \right\} - i \pi \delta(\omega - \omega')\,.
\end{equation}
It is important to remark that, for $V_0=0$ the identity (\ref{prop}) reduces to the
following identity
between equilibrium Green's functions:
\begin{equation}\label{propeq}
\hat{\rho}(\omega) \equiv \hat{G}^{0,R}(\omega)-\hat{G}^{0,R\dagger}(\omega)= -i 
\hat{G}^{0,R}(\omega)\hat{\Gamma}(\omega) 
\hat{ G}^{0,R}(\omega)^{\dagger}\,.
\end{equation}

\subsubsection{Perturbative solution of the Dyson's equation}  \label{per}
Sometimes, in order to derive analytical expressions, it is convenient to solve the  set
(\ref{gretw}) 
by recourse to a perturbative expansion in $\hat{V}$ (see Arrachea 2005). The solution up to second order
in this parameter is obtained by writing (\ref{gretw}) evaluated at 
$\omega_n$, for $n=-2,\ldots,2$ and back-substituting the equation evaluated at
$\omega_2$ into the one evaluated at $\omega_1$, and the latter into the one 
evaluated at $\omega$, and a similar procedure with  $\omega_{-2} \rightarrow \omega_{-1}$
and the latter into $\omega$. If we then collect all the coefficients of 
 $e^{-i n \Omega_{0} t}$
in the resulting expression and recalling the representation (\ref{gfou}), we obtain
\begin{equation}\label{gfou2}
\hat{G}^R (t,\omega) \sim \sum_{n=-2}^{+2} {\cal G}(n,\omega) e^{-i n \Omega_{0} t}\,,
\end{equation}
with:
\begin{eqnarray}\label{gper}
\hat{\cal G}(0,\omega)&=& \hat{G}^{0,R}(\omega)+ \sum_{k=\pm 1}\hat{\cal G}( k,\omega)\hat{V}(-k)\,,
\nonumber \\
\hat{\cal G}(\pm 1,\omega)&=&\hat{G}^{0,R}(\omega_{ \pm 1})\hat{V}(\pm 1) \hat{G}^{0,R}(\omega)\,,
 \nonumber \\
\hat{\cal G}(\pm 2,\omega)&=&\hat{G}^{0,R}(\omega_{\pm 2})\hat{V}(\pm 1)\hat{\cal G}(\pm 1,\omega)\,.
\end{eqnarray} 
The reader can easily extend the procedure to evaluate higher order terms.

\subsection{Scattering matrix formalism}

%As we already mentioned, 
To calculate the charge and energy flows generated by the driven central system in the wires one can also use the scattering approach.
Within this approach we consider the central system as some scatterer which reflects or transmits electrons incoming from the wires. 
The electrons coming, for instance, from the left wire can be transmitted to the right wire or can be reflected back to the left wire. 
To find the current in some wire we need just to calculate the difference between the number of particles incoming through this wire and the number of particles exiting the central system through the same wire.
We do not need to know what happened with an electron inside the central system. 
We only need to know the quantum-mechanical scattering amplitudes for an electron to be transmitted/reflected through/from the central system. 
The advantage of the scattering approach is the simplicity and the physical transparency of expressions written in terms of scattering amplitudes. 
We stress that the scattering approach does not aim to calculate the single particle scattering amplitudes. 
This approach tells us how to calculate the transport properties of a mesoscopic structure coupled to wires if the scattering amplitudes are known. 
To calculate the scattering amplitudes one can use the Green's functions method. 
We will give an explicit expression for the scattering amplitudes in terms of corresponding Green's functions.
Actually the combining Green's functions -- scattering approach is one of the most powerful and practical approaches for transport phenomena in mesoscopic structures.

\subsubsection{General formalism}

The scattering approach to transport phenomena in small phase-coherent samples connected to macroscopic reservoirs was introduced and developed by Landauer and B\"uttiker (Landauer 1957, 1970, 1975, B\"uttiker 1990, 1992, 1993). 

Within this formalism we consider electrons only in the one-dimensional wires connecting the central system to macroscopic reservoirs. 
It is convenient to introduce separate operators $a_{\alpha}(\varepsilon)$ for incoming and $b_{\alpha}(\varepsilon)$ for scattered electrons with
energy $\varepsilon$. 
 
Then the current $J_{\alpha}(t)$ flowing into wire $\alpha$ to the central system is the following (B\"uttiker 1992):
\begin{equation}
\label{smf_1}
J_{\alpha}(t) =  \frac{e}{h} \int_{-\infty}^{\infty} d\varepsilon \, d\varepsilon^{\prime}
e^{i \frac{\varepsilon-\varepsilon^{\prime}}{\hbar} t }  \left\{
\langle  a^{\dagger}_{\alpha}(\varepsilon) a_{\alpha}(\varepsilon^{\prime})\rangle 
- \langle  b^{\dagger}_{\alpha}(\varepsilon) b_{\alpha}(\varepsilon^{\prime})\rangle 
\right\}.
\end{equation}
Here $\langle \dots \rangle$ denotes averaging over equilibrium states of reservoirs. 

Correspondingly the dc current reads:
\begin{equation}
\label{smf_01}
\overline J_{\alpha} =  \frac{e}{h} \int_{-\infty}^{\infty} d\varepsilon \, \left\{ f_{\alpha}(\varepsilon) - f^{(out)}_{\alpha}(\varepsilon) \right\},
\end{equation}
where $ f^{(out)}_{\alpha}(\varepsilon) = \langle  b^{\dagger}_{\alpha}(\varepsilon) b_{\alpha}(\varepsilon)\rangle $ is the distribution function for electrons exiting the central system through the wire $\alpha$, and $f_{\alpha}(\varepsilon) = \langle  a^{\dagger}_{\alpha}(\varepsilon) a_{\alpha}(\varepsilon)\rangle $ is the distribution function for electrons incoming through the wire $\alpha$. 
This expression tells us that the dc current is the difference per unit time
between the number of electrons entering and exiting 
the system. 
As the reservoir is at equilibrium, the distribution function for the
incoming electrons is 
the Fermi distribution function. 
In contrast, the scattered electrons, in general, are non equilibrium particles. 
To calculate the distribution function for the scattered electrons we express the $b$-operators in terms of $a$-operators.
Since an electron coming from any wire can be scattered into a given
 wire $\alpha$, then the operators $b_{\alpha}$ depend on all the operators  for the incoming particles.
In the model we consider in this chapter, the number of reservoirs is two, $\beta = L,\,R$.
%$N_{r} = 2$. 
Therefore,  $b_{\alpha}(\varepsilon) = \sum_{\beta=L,R}^{} S_{\alpha\beta}(\varepsilon) a_{\beta}(\varepsilon)$, being $S_{\alpha\beta}$
the scattering amplitudes. 
These amplitudes are normalized in such a way that their square define corresponding currents (B\"uttiker 1992).
The quantities $S_{\alpha\beta}(\varepsilon)$ can be viewed as the elements of some matrix, which is called the scattering matrix $\hat S(\varepsilon)$.

\subsubsection{Floquet scattering matrix}

If the scatterer is driven by external forces which are periodic in time with period $\tau = 2\pi/\Omega_{0}$, then interacting with such a scatterer an electron can gain or loss some energy quanta $n\hbar\Omega_{0}$, $n=0,\pm 1,\dots$.
Therefore, in this case the scattering amplitudes in addition to the two wire indexes become dependent on the two energies, one for the incoming and the other
for the outgoing electrons.
Such a scattering matrix is called the Floquet scattering matrix $\hat S_{F}$ (see, e.g., Platero and Aguado 2004).  
Their elements, $S_{F,\alpha\beta}(\varepsilon_n, \varepsilon)$, are related
to photon-assisted amplitudes  for an electron  with energy $\varepsilon$ 
entering the scatterer through the lead $\beta$ and leaving
 the scatterer with energy  $\varepsilon_n = \varepsilon + n\hbar\Omega_{0}$ 
through the lead $\alpha$. 
Now the relation between the operators $b$ for outgoing particles and $a$ for incoming particles reads (Moskalets and B\"uttiker 2002a): 
\begin{equation}
\label{smf_3}
b_{\alpha}(\varepsilon) = \sum\limits^{}_{\beta=L,R}\sum\limits_{n} S_{F,\alpha\beta}(\varepsilon, \varepsilon_{n}) a_{\beta}(\varepsilon_n)\,,
\end{equation}
where the sum over $n$ runs over those $n$ for which $\varepsilon_{n}  > \varepsilon_{0\beta}$, hence it corresponds to propagating (i.e. current-carrying) states. 
We denote the Floquet scattering matrix  the submatrix corresponding to transitions between the propagating states only. 
In the case where $\hbar \Omega_{0} \ll \varepsilon$, the sum in Eq.\,(\ref{smf_3}) runs over all the integers: $-\infty < n < \infty$.  
In what follows we assume this to be the case.
Note that if the scatterer is stationary,  the only term that remains non-vanishing is that with  $n=0$, and the Floquet scattering matrix is reduced to the stationary scattering matrix with elements $S_{\alpha\beta}(\varepsilon) = S_{F,\alpha\beta}(\varepsilon, \varepsilon)$.

The conservation of the particle current at each scattering event implies that the Floquet scattering matrix is a unitary matrix (Moskalets and B\"uttiker 2002a, 2004): 
\begin{equation}
\label{smf_4}
\sum\limits_{\alpha = L,R}^{} \sum\limits_{ n=-\infty }^{ \infty}  S^{*}_{F, \alpha\beta}( \varepsilon_n, \varepsilon )\, S_{F, \alpha\gamma}( \varepsilon_n, \varepsilon_m) = \delta_{m0}\, \delta_{\beta \gamma}\,, 
\end{equation}
\begin{equation}
\label{smf_4_1}
\sum\limits_{\beta = L,R }^{} \sum\limits_{ n=-\infty }^{ \infty} S^{*}_{F, \alpha\beta }(\varepsilon, \varepsilon_n) \, S_{F, \gamma\beta}(\varepsilon_m, \varepsilon_n) = \delta_{m0} \, \delta_{\alpha\gamma}\,.
\end{equation}

Using Eq.\,(\ref{smf_3}) we calculate the distribution function for electrons scattered into wire $\alpha$:
\begin{equation}
\label{smf_6}
f^{(out)}_{\alpha}(\varepsilon) = \sum\limits_{\beta = L,R}^{} \sum\limits_{n = - \infty}^{\infty} \left|S_{F,\alpha\beta}(\varepsilon, \varepsilon_{n}) \right|^2 f_{\beta}(\varepsilon_{n})\,.
\end{equation}
This function is not the Fermi distribution function unless the scatterer is stationary and all the reservoirs have the same chemical potentials and temperatures. 
This reflects the fact that the particles scattered by the dynamical scatterer (quantum pump) are out of equilibrium. 

\subsubsection{Adiabatic scattering}\label{adiabatic}

If the driving forces change slowly, $\Omega_{0} \to 0$, they behave as if they were almost constant for
the electrons propagating through the central system.
For this reason, the scattering properties of a slowly driven  (adiabatic) scatterer are close to those of a stationary one. 
Nevertheless there is an essential difference: in spite of the slowly change of
the fields, 
 an electron can still  absorb or emit one or several energy quanta 
$\hbar\Omega_{0}$ in its travel through the central system.  
Therefore, although the adiabatic scatterer is characterized by the Floquet scattering matrix dependent on two energies, $\hat S_{F}(\varepsilon_{n}, \varepsilon)$, it is natural to expect that it could be related to
 the stationary scattering matrix  $\hat S(\varepsilon)$ under these conditions

The stationary scattering matrix $\hat S$ depends on the electron energy $\varepsilon$ and some properties of the scatterer. 
To account the latter dependence we introduce the set of parameters, $\{p_{i} \}$, $i = 1, \dots, M_{p}$ and write $\hat S(\{p_{i} \}, \varepsilon)$. 
Under the action of external periodic forces the parameters periodically change in time, $p_{i}(t) = p_{i}(t + \tau ) $. 
Therefore, the matrix $\hat S$ becomes dependent on time, $\hat S(t, \varepsilon) \equiv \hat S(\{p_{i}(t) \}, \varepsilon)$ and periodic, $\hat S(t, \varepsilon) = \hat S(t + \tau, \varepsilon)$. 
The obtained matrix is called the frozen scattering matrix.
This name means that the matrix $\hat S(t_{0}, \varepsilon)$ describes the scattering properties of  a stationary scatterer whose parameters coincide with the parameters of a given scatterer frozen at time $t  = t_{0}$.  
The Fourier coefficient for the frozen matrix, 
\begin{equation}
\hat S_{n}(\varepsilon) = \int_{0}^{\tau} \frac{dt}{\tau} \, e^{in\Omega_{0} t }\, \hat S(t,\varepsilon)\,,
\label{froz}
\end{equation}
 can be related to the Floquet scattering matrix.

At low driving frequencies, $\Omega_{0}\to 0$, one can expand the elements of the Floquet scattering matrix in powers of $\Omega_{0}$. 
Up to the first order in $\Omega_{0} $ we have (Moskalets and B\"uttiker 2004):
\begin{equation}
\label{smf_17}
\hat S_{F}(\varepsilon_n, \varepsilon)  = \hat S_{n}(\varepsilon) + \frac{ n \hbar \Omega_{0}}{2} \frac{ \partial \hat S_{n}(\varepsilon) }{ \partial \varepsilon}  + \hbar \Omega_{0} \hat A_{n}(\varepsilon) + {\cal O}(\Omega_{0}^2) \,.
\end{equation}
Here $\hat A_{n}$ is the Fourier transform for a matrix 
$\hat A(t, \varepsilon)$, which formally encloses corrections that can not be related to the frozen scattering matrix and has to be calculated independently, see (Moskalets and B\"uttiker 2005) for some examples. 
Note that in the above equation the frozen scattering matrix and the matrix $\hat A$ should be kept as energy-independent within a scale of order $\hbar\Omega_{0}$.
 
The unitarity of the Floquet scattering matrix puts some constraint on the matrix $\hat A$. 
Substituting Eq.\,(\ref{smf_17}) into Eq.\,(\ref{smf_4}) and taking into account that the stationary (frozen) scattering matrix is unitary we get the 
following relation:
\begin{equation}
\label{smf_18}
\hbar\Omega_{0}\Big\{ \hat S^{\dagger}\hat A + \hat A^{\dagger}\hat S\Big\} = \frac{i\hbar}{2}\left( \frac{\partial\hat S^{\dagger}}{\partial t} \frac{\partial\hat S}{\partial \varepsilon} - \frac{\partial\hat S^{\dagger}}{\partial \varepsilon} \frac{\partial\hat S }{\partial t}\right)\,.
\end{equation}

The advantage of the adiabatic ansatz, Eq.\,(\ref{smf_17}), is that the matrices $\hat S$ and $\hat A$ depend only on one energy and thus have a much smaller number of 
elements than the Floquet scattering matrix. 
In addition, the adiabatic ansatz allows us to draw  some conclusions concerning the physical properties of slowly driven systems, in particular, concerning the generated 
heat flows.

\subsection{Floquet scattering matrix versus Green's function}
There exists a simple relation between the Floquet scattering matrix elements and the Fourier coefficients for the Green's function (Arrachea and Moskalets 2006):
\begin{equation}
\label{smf_24}
S_{F,\alpha \beta}(\hbar\omega_m, \hbar\omega_n)= \delta_{\alpha,\beta}\, \delta_{m,n} - i  \sqrt{ \Gamma_{\alpha}( \omega_{m})
 \Gamma_{\beta}(\omega_{n}) }\, { \cal G}_{l \alpha, l \beta }(m-n,\omega_{n})\,,
\end{equation}
where the Floquet component of the Fourier transformed Green's function ${\cal G}(n,\omega)$ was introduced in Eq.\,(\ref{gfou}).   
The equation (\ref{smf_24}) is a generalization to  periodically driven systems of a formula proposed by
 Fisher and Lee (Fisher and Lee 1981) for stationary systems. This relation is based in the fact that the unitary property
(\ref{smf_4}) and (\ref{smf_4_1}) which is  fundamental to prove the conservation of the charge within the scattering matrix formalism
can be proved from identities between the Green's functions, see Eq.\,(\ref{prop}) through the relation (\ref{smf_24}).
We do not present in this chapter further details on those proofs. Instead, in the next subsection we explicitly show that both
formalisms lead to expressions for the currents through the contacts that are equivalent provided that the above relation holds.

\subsection{Final expressions for the dc currents and powers}

\subsubsection{Particle currents and the conservation of the charge}
We begin with the expression for the dc particle currents within the Green's function formalism.
In the subsection (\ref{exp}) we have expressed instantaneous values of observables in terms of 
lesser Green's functions. Now, we use those expressions to evaluate the 
dc components of the currents defined in subsection
(\ref{curdef}). In particular, for the charge currents (\ref{jl}) and (\ref{jal}) we have:
\begin{eqnarray}
\overline{J}_l &=& \frac{2 e w }{\hbar \tau }  \int_0^{\tau}
dt   \mbox{Re}[ G^<_{l+1,l}(t,t)]\,, \nonumber \\
\overline{J}_{\alpha} &=& \frac{2 e w_{c\alpha} }{\hbar \tau}   \int_0^{\tau}
dt  \mbox{Re}[ G^<_{l\alpha,k\alpha}(t,t)]\,.
\end{eqnarray}
Using the representation  (\ref{gfou}) in (\ref{glesc}) and (\ref{glesin}) 
and substituting in the above expressions
casts for the charge currents within $C$:
\begin{eqnarray}
\overline{J}_l & = & -  \frac{2 e w }{h} \sum_{\alpha=L,R} \sum_{n=-\infty}^{\infty}
\int_{-\infty}^{+\infty} {d \omega}
f_{\alpha}(\hbar\omega)\Gamma_{\alpha}(\omega)
\mbox{Im}[{\cal G}_{l+1,l\alpha}(n,\omega) {\cal G}^*_{l,l\alpha}(n,\omega)]\,,
\end{eqnarray}
and through the contacts
\begin{eqnarray} 
\overline{J}_{\alpha} & = & - \frac{2 e |w_{c \alpha}|^2 }{h}
\int_{-\infty}^{+\infty} {d \omega}
\mbox{Re}\bigg\{i  f_{\alpha}(\hbar\omega) {\cal G}_{l\alpha,l\alpha}(0, \omega )
\rho_{\alpha}(\omega) \nonumber \\
& & +  \sum_{\beta=L,R} \sum_{n=-\infty}^{+\infty} \sum_{k \alpha}
 f_{\beta}(\hbar\omega) |{\cal G}_{l\alpha,l\beta}(n, \omega )|^2
\Gamma_{\beta}(\omega) g^{0,A}_{k \alpha,k \alpha }(\omega_n)\bigg\} \nonumber \\
 & = &  \frac{ e }{h}
\int_{-\infty}^{+\infty} {d \omega} 
 \bigg\{ f_{\alpha}(\hbar\omega) \Gamma_{\alpha}(\omega) 2 \mbox{Im}[{\cal G}_{l\alpha,l\alpha}(0, \omega )]
\nonumber \\
& & -  \sum_{\beta=L,R} \sum_{n=-\infty}^{+\infty} 
 f_{\beta}(\hbar\omega) \Gamma_{\alpha}(\omega_n) |{\cal G}_{l\alpha,l\beta}(n, \omega )|^2
\Gamma_{\beta}(\omega) \bigg\}\,.\label{part1}
\end{eqnarray}
In the above equations 
 we have used the definitions of the density of states (\ref{dos}) and the functions 
(\ref{sigmas}). Going from the first to the second identity, we
have also used the property  $\mbox{Im}[g^{0,R}_{k \alpha,k \alpha }(\omega)]=
-\mbox{Im}[g^{0,A}_{k \alpha,k \alpha }(\omega)]=-\rho_{\alpha}(\omega)/2$, which can be 
easily derived just evaluating the Fourier transforms in (\ref{gres}).
From the identity (\ref{prop}), this current can also be expressed in the more compact and symmetric form:
\begin{eqnarray} \label{partsim}
\overline{J}_{\alpha} & = &  \frac{ e }{h }
\sum_{\beta=L,R} \sum_{n=-\infty}^{+\infty}
\int_{-\infty}^{+\infty} {d \omega}[f_{\alpha}(\hbar\omega_n)-f_{\beta}(\hbar\omega)]
\Gamma_{\alpha}(\omega_n) |{\cal G}_{l\alpha,l\beta}(n, \omega )|^2 \Gamma_{\beta}(\omega)\,.
\end{eqnarray}

Within the Floquet scattering matrix approach, we proceed as follows.
Substituting Eq.\,(\ref{smf_6}) into Eq.\,(\ref{smf_01}) we get the current in terms of the Floquet scattering matrix elements:
\begin{equation}
\label{smf_9}
\overline J_{\alpha} = \frac{e}{h}  \int_{-\infty}^{+\infty} d\varepsilon  \left\{  f_{\alpha}(\varepsilon) - \sum\limits_{ \beta=L,R } \sum\limits_{ n = - \infty }^{\infty} f_{\beta}(\varepsilon_n)  \left|S_{F,\alpha\beta}(\varepsilon, \varepsilon_n) \right|^2  \right\}.
\end{equation} 
An equivalent expression is obtained if  
we make a shift $\varepsilon_{n} \to \varepsilon$ (under the integration over energy) and an inversion $n \to -n$ (under the corresponding sum) in the term containing $f_{\beta}(\varepsilon_{n})$. The results is:
\begin{equation}
\label{smf_10}
\overline J_{\alpha} = \frac{e}{h}  \int_{-\infty}^{+\infty} d\varepsilon  \left\{ f_{\alpha}(\varepsilon) - \sum\limits_{ \beta=L,R } \sum\limits_{ n = - \infty }^{\infty} f_{\beta}(\varepsilon)  \left|S_{F,\alpha\beta}(\varepsilon_{n}, \varepsilon) \right|^2  \right\}.
\end{equation}
Finally, we can write this equation in an alternative way as follows.
We multiply
 the term $f_{\alpha}(\varepsilon)$ in Eq.\,(\ref{smf_10}) by the left hand side of the identity, 
$\sum_{\beta}\sum_{n} \left| S_{\alpha\beta}(\varepsilon, \varepsilon_{n}) \right|^2 = 1$,  
following from the unitarity condition Eq.\,(\ref{smf_4_1}), change  $\varepsilon_{n} \to \varepsilon$ and $n\to -n$ in the resulting expression, and find:
\begin{equation}
\label{smf_8}
\overline J_{\alpha} = \frac{e}{h} \sum\limits_{ \beta = L,R } \sum\limits_{ n = - \infty}^{\infty} \int_{-\infty}^{+\infty} d\varepsilon \left[ f_{\alpha}(\varepsilon_{n}) - f_{\beta}(\varepsilon) \right] \left|S_{F,\alpha\beta}(\varepsilon_{n},\varepsilon) \right|^2  \,.
\end{equation}

It is  important to note that Eq.\,(\ref{smf_9}) coincides with  (\ref{part1}), while 
(\ref{smf_8}) coincides with (\ref{partsim})
if we apply the relation (\ref{smf_24}) between
the Floquet scattering matrix and the Green's function.

Another feature worth of being mentioned is the fact that from the expressions (\ref{partsim}) and (\ref{smf_8}) it can be proved 
 the conservation of the charge, which implies:
\begin{equation}
\label{smf_9_1}
\sum\limits_{\alpha=L,R} \overline J_{\alpha} = 0\,.
\end{equation}
We recall that the $\overline{J}_{\alpha}$ was defined as the current exiting the reservoir, for this reason current conservation implies
that it has different signs at the two reservoirs. A final issue that becomes apparent from Eqs. (\ref{partsim}) and (\ref{smf_8}),
is the fact that for slow driving, $\Omega_{0}\to 0$, only electrons near the Fermi energy  $\varepsilon \approx \mu$
will be excited and hence will contribute to 
the generated current, in agreement with our intuition. 

\subsubsection{Particle currents within the adiabatic approximation}
In the subsection (\ref{adiabatic}) we have introduced an
approximation for the low driving limit of the full Floquet scattering matrix
that depends on the frozen scattering matrix and a matrix  $\hat A$. In this section we present the expression for the
current in terms of that approximation.

We have mentioned that the unitary condition imposes a constraint to the matrix $\hat A$.  
Another more specific constraint  follows from the conservation of a charge current expressed directly in terms of $\hat S$ and $\hat A$ matrices.
To derive it we calculate the dc pumped current $\overline J_{\alpha}$ up to $\Omega_{0}^2$ terms for all reservoirs at the same temperature and
chemical potential, i.e. $f_{\alpha} = f_{0}, \forall \alpha$. 
Since in the adiabatic case under consideration $\Omega_{0} \to 0$, then at any finite temperature it is $k_{B} T \gg \hbar\Omega_{0}$, and we can expand $f_{0}(\varepsilon) -  f_{0}(\varepsilon_{n}) \approx -(\partial f_{0}/\partial\varepsilon) n\hbar\Omega_{0} - (\partial^{2} f_{0}/\partial\varepsilon^{2}) (n\hbar\Omega_{0})^2/2$.
Substituting this expansion and Eq.\,(\ref{smf_17}) into Eq.\,(\ref{smf_8})
and performing the inverse Fourier transformation we calculate the charge current as a sum of linear (upper index ``(1)'' ) and quadratic (upper index ``(2)'' ) in driving frequency contributions, $\overline J_{\alpha} = \overline J^{(1)}_{\alpha} + \overline J^{(2)}_{\alpha} + {\cal O}\left(\Omega_{0}^{3} \right)$, with
\begin{equation}
\label{smf_18_1}
\overline J^{(1)}_{\alpha} =  -\, \frac{e}{2\pi} \int_{-\infty}^{\infty} d\varepsilon \,\left( - \frac{\partial f_{0} }{\partial \varepsilon } \right) \int_{0}^{\tau} \frac{dt}{\tau} \, 
\mbox{Im} \left( \hat S(t,\varepsilon) \frac{\partial \hat S^{\dagger}(t,\varepsilon) }{\partial t} \right)_{\alpha\alpha} \,, \\
\end{equation}
\begin{equation}
\label{smf_18_2}
\overline J^{(2)}_{\alpha} = -\, \frac{e}{2\pi} \int_{-\infty}^{\infty} d\varepsilon \, \left( - \frac{\partial f_{0} }{\partial \varepsilon } \right) \int_{0}^{\tau} \frac{dt}{\tau}\, \mbox{Im} \left( 2\Omega_{0} \hat A(t,\varepsilon)  \frac{\partial \hat S^{\dagger}(t,\varepsilon) }{\partial t} \right)_{\alpha\alpha}  \,. 
\end{equation}

The linear behavior of the current as a function of the frequency  was calculated by Brouwer (Brouwer 1998) using the scattering approach to low-frequency ac transport in mesoscopic systems developed by B\"uttiker et al. (B\"uttiker et al. 1994). 
The conservation of this current, $\sum_{\alpha} \overline J^{(1)}_{\alpha} = 0$, was demonstrated by Avron et al. (Avron et al. 2004) on the base of the Birman-Krein relation, $d \ln (\det\hat S) = - {\rm Tr} (\hat S d\hat S^{\dagger})$  (where $\det (\hat X)$ and ${\rm Tr} (\hat X)$ are the determinant and the trace of a 
matrix $\hat X$, respectively), applied to the frozen matrix which is unitary. 

The conservation of the current up to the second order in frequency, $\sum_{\alpha} \overline J^{(2)}_{\alpha} = 0$, leads to the constraint for the matrix $\hat A$ we are looking for: 
\begin{equation}
\label{smf_18_3}
\mbox{Im} \, \int_{0}^{\tau} \frac{dt}{\tau} \, {\rm Tr} \, \left( \hat A \, \frac{\partial \hat S^{\dagger} }{\partial t} \right)  = 0\,,
\end{equation}

Equations (\ref{smf_17}) and (\ref{smf_18}) show us that the expansion in powers of $\Omega_{0}$ 
actually is an expansion in powers of $\hbar\Omega_{0}/\delta \varepsilon$, where $\delta \varepsilon$ is an energy scale characteristic for the stationary scattering matrix. 
The energy $\delta \varepsilon$ relates to the inverse time spent by an electron with energy $\varepsilon$ inside the scattering region (the dwell time).
Therefore, one can say that  the   adiabatic expansion, Eq.\,(\ref{smf_17}) is valid if the period of external forces is large compared with the dwell time.
It is important to stress that this definition of  ``adiabaticity'' is different from that
usually used in quantum mechanics one which requires the excitation quantum $\hbar\Omega_{0}$ to be 
small compared with the level spacing.

\subsubsection{Particle currents within perturbation theory}\label{curpert}
In order to gain physical intuition on the behavior of the dc charge current, let us consider the 
weak  driving regime (low $V_0$) and let us evaluate  (\ref{partsim}) $\overline{J}_{\alpha}$
with the perturbative solution
of the Green's function we have presented in (\ref{gfou2}). We assume that both reservoirs are at temperature $T_{\alpha}=0$.
Substituting (\ref{gfou2}) into (\ref{partsim}), we get:
\begin{equation}
\overline{J}_{\alpha} = \frac{e}{h} \sum_{\beta=L,R}\sum_{k=\pm 1} 
\int_{-\infty}^{\infty} d\omega [f_{\alpha}(\hbar\omega_k)-f_{\beta}(\hbar\omega)] \Gamma_{\alpha}(\omega_k) |{\cal G}_{l_{\alpha},l_{\beta}}(k,\omega)|^2 \Gamma_{\beta}(\omega)\,.
\end{equation}
In the same spirit as in the adiabatic approximation, let us consider that the driving is slow, i.e. $\Omega_{0} \rightarrow 0$, and let us expand the 
integrand of the above equation up to the first order in $\Omega_{0}$. Replacing the Floquet components evaluated up to second order in
perturbation theory (\ref{gper}) we get:
\begin{eqnarray}\label{jpertad}
\overline{J}_{\alpha} & = & \frac{2eV_0^2 \Omega_{0}}{h} \sum_{j,j'=1}^M \sum_{\beta=L,R}  \Gamma_{\alpha}(\mu)  \Gamma_{\beta}(\mu) 
\sin(\delta_j-\delta_{j'})   G^{0,R}_{l\alpha,lj}(\mu) G^{0,R}_{lj,l\beta}(\mu) \left[G^{0,R}_{l\alpha,lj'}(\mu) G^{0,R}_{lj',l\beta}(\mu)\right]^*\,, \nonumber \\
%& = & - \frac{4 V_0^2 \Omega_{0}}{h} \sum_{j,j'=1}^M \Gamma_{\alpha}(\mu) 
%\sim(\delta_j-\delta_{j'}) \mbox{Im}\{ G^{0,R}_{l\alpha} lj}(\mu) [G^{0,R}_{l\alpha} lj'}(\mu)]^*}
%Im[G^{0,R}_{lj lj' }(\mu)],
\end{eqnarray}
where $j,j'$ runs over the $M$ pumping potentials.
Thus, even without specifying the geometrical details on the structure, which
 are contained in $G^0$,
Eq.\,(\ref{jpertad}) provides us a valuable piece of information. 
As a first point 
it tells us that at low driving
the leading contribution to the dc particle current  is $\propto V_0^2 \Omega_{0}$
A second important point is that 
 with local time-dependent potentials, as we are considering in our model,  we need 
at least two of these potentials operating with a phase lag in order to have a non-vanishing value for this
lowest order contribution.

\subsubsection{Energy and heat currents}
We can follow a similar procedure as in the previous subsection to derive the dc energy and heat currents.
In terms of Green's functions we start writing  the 
dc energy currents (\ref{jel}) and (\ref{jeal}) as follows:
\begin{eqnarray}
\overline{J}^E_l &=& \frac{2  w }{\hbar \tau}   \int_0^{\tau}
dt  \left\{\mbox{Re}[ G^<_{l+2,l}(t,t)] w - 
\mbox{Re}[ G^<_{l+2,l+1}(t,t)]\varepsilon_{l+1}(t) \right\} , 
 \nonumber \\
\overline{J}^E_{\alpha} &=& \frac{2  w_{c\alpha} }{\hbar \tau}  \sum_{k \alpha} \int_0^{\tau}
dt\, \varepsilon_{k \alpha}\, \mbox{Re}[ G^<_{l\alpha,k\alpha}(t,t)]\,,
\end{eqnarray}

The energy  current within $C$ is:
\begin{eqnarray}
\overline{J}^E_l & = & -  \frac{2  w }{h } \sum_{\alpha=L,R} \sum_{n=-\infty}^{\infty}
\int_{-\infty}^{+\infty} {d \omega}
f_{\alpha}(\hbar\omega)\Gamma_{\alpha}(\omega) \big\{ w
\mbox{Im}[{\cal G}_{l+2,l\alpha}(n,\omega) {\cal G}^*_{l,l\alpha}(n,\omega)] \nonumber \\
& & 
-\varepsilon_{l+1}\mbox{Im}[{\cal G}_{l+2,l\alpha}(n,\omega) {\cal G}^*_{l+1,l\alpha}(n,\omega)] \big\}\,,
\end{eqnarray}
where we have assumed that the position $l+1$ does not coincide with a pumping center,
while for the energy current through the contact we get:
\begin{eqnarray} 
\overline{J}^E_{\alpha} & = &  \frac{  |w_{c \alpha}|^2 }{h }
\sum_{k \alpha} \int_{-\infty}^{+\infty} {d \omega}\, \varepsilon_{k \alpha}\,
2 \pi 
 \bigg\{ f_{\alpha}(\hbar\omega) \delta(\omega- \varepsilon_{k \alpha}/\hbar)2 \mbox{Im}[{\cal G}_{l\alpha,l\alpha}(0, \omega )]
\nonumber \\
& & -  \sum_{\beta=L,R} \sum_{n=-\infty}^{+\infty} 
 f_{\beta}(\hbar\omega) \delta(\omega_n - \varepsilon_{k \alpha}/\hbar)|{\cal G}_{l\alpha,l\beta}(n, \omega )|^2
\Gamma_{\beta}(\omega) \bigg\}\nonumber \\
 & = &  \frac{ \hbar }{2\pi}
 \int_{-\infty}^{+\infty} {d \omega}
 \bigg\{  \omega  f_{\alpha}(\hbar\omega)\Gamma_{\alpha}(\omega) 2 \mbox{Im}[{\cal G}_{l\alpha,l\alpha}(0, \omega )]
\nonumber \\
& & -  \sum_{\beta=L,R} \sum_{n=-\infty}^{+\infty} \omega_n
 f_{\beta}(\hbar\omega)  \Gamma_{\alpha}(\omega) |{\cal G}_{l\alpha,l\beta}(n, \omega )|^2
\Gamma_{\beta}(\omega) \bigg\}\,, \label{jens}
\end{eqnarray}
which can also be written in the symmetric form:
\begin{eqnarray} \label{sime}
\overline{J}^E_{\alpha} & = &  \frac{ \hbar }{2\pi }
\sum_{n=-\infty}^{+\infty}\sum_{\beta=L,R}
\int_{-\infty}^{+\infty} {d \omega}\,\omega_n\, [f_{\alpha}(\hbar\omega_n) -f_{\beta}(\hbar\omega)]
 \Gamma_{\alpha}(\omega_n) |{\cal G}_{l\alpha,l\beta}(n, \omega )|^2\, \Gamma_{\beta}(\omega)\,.
\end{eqnarray}
We now go back to our heuristic argument introduced in Section \ref{heat} to define the heat current.
The above equation shows that for low driving, even for reservoirs at $T=0$ and very weak driving, such that $\Omega_{0} \rightarrow 0$, there is a finite energy flow  $\overline{J}^E_{\alpha} \propto \mu \overline{J}_{\alpha}$, with $\overline{J}_{\alpha}$ given in (\ref{partsim}). 
This energy is transported by the currents from one reservoir to the other one, thus having a convective character and should be subtracted to get a heat flow.

To calculate the heat flow we multiply (\ref{part1}) by $\mu/e$ and subtract it to  (\ref{jens}):
\begin{eqnarray} \label{jq1}
J^Q_{\alpha} & = &    \frac{ \hbar }{2\pi}
 \int_{-\infty}^{+\infty} {d \omega} 
\bigg\{ (\omega- \omega_{F}) f_{\alpha}(\hbar\omega)\Gamma_{\alpha}(\omega)  2 \mbox{Im}[{\cal G}_{l\alpha,l\alpha}(0, \omega )]
\nonumber \\
& & -  \sum_{\beta=L,R} \sum_{n=-\infty}^{+\infty} 
(\omega_n- \omega_{F}) 
f_{\beta}(\hbar\omega) \Gamma_{\alpha}(\omega_n) |{\cal G}_{l\alpha,l\beta}(n, \omega )|^2
\Gamma_{\beta}(\omega) \bigg\}\,, \label{jqal}
\end{eqnarray}
where $\hbar\omega_{F} = \mu$.
Equivalently, from (\ref{sime}) and (\ref{partsim}), we can write the heat current flowing through the contact as follows:
\begin{eqnarray} \label{jqsim}
J^Q_{\alpha} & = &  \frac{ \hbar }{2\pi }
\sum_{\beta=L,R} \sum_{n=-\infty}^{+\infty} 
\int_{-\infty}^{+\infty} {d \omega}(\omega_n -\omega_{F})
[f_{\alpha}(\hbar\omega_n)-f_{\beta}(\hbar\omega)]
 \Gamma_{\alpha}(\omega_n) |{\cal G}_{l\alpha,l\beta}(n, \omega )|^2 \Gamma_{\beta}(\omega)\,. \nonumber \\
\end{eqnarray}

Within the scattering matrix approach,  one can also calculate the heat current $J^{Q}_{\alpha}$
by analogy to the charge current, Eq.\,(\ref{smf_9}). 
As we already mentioned, Eq.\,(\ref{smf_9}) contains the difference of number of electrons with energy $\varepsilon$ entering and leaving the scatterer 
through the same wire. Each of these electrons has an energy $\varepsilon$. 
Therefore, to calculate the heat current we multiply the integrand in Eq.\,(\ref{smf_9}) by $(\varepsilon - \mu)$, drop an electron charge $e$, and get:
\begin{equation}
\label{smf_11}
J^{Q}_{\alpha} = \frac{1}{h} \int_{-\infty}^{+\infty} d\varepsilon \, (\varepsilon - \mu) \left\{ f_{\alpha}(\varepsilon) -  \sum\limits_{ \beta=L,R } \sum\limits_{ n = - \infty }^{\infty}  f_{\beta}(\varepsilon_n) \left|S_{F,\alpha\beta}(\varepsilon, \varepsilon_n) \right|^2    \right\}.
\end{equation}
This equation is equivalent to (\ref{jq1}) through the relation (\ref{smf_24}). 
Next we make shifts $\varepsilon_{n} \to \varepsilon$ and $n \to -n$ in the term containing $f_{\beta}(\varepsilon_{n})$ and finally obtain:
\begin{equation}
\label{smf_12}
J^{Q}_{\alpha} = \frac{1}{h} \int_{-\infty}^{+\infty} d\varepsilon \left\{  (\varepsilon-\mu) f_{\alpha}(\varepsilon) - \sum\limits_{ \beta=L,R} \sum\limits_{ n = - \infty }^{\infty} (\varepsilon_{n} - \mu) \, f_{\beta}(\varepsilon) \left|S_{F,\alpha\beta}(\varepsilon_{n}, \varepsilon) \right|^2  \right\},
\end{equation}
We multiply the term containing $f_{\alpha}(\varepsilon)$ by the identity $1 = \sum_{\beta}\sum_{n} | S_{F,\alpha\beta} (\varepsilon, \varepsilon_{n}) |^2$.
Then we make shifts $\varepsilon_{n} \to \varepsilon$ and $n \to -n$ in this term and finally get:
\begin{equation}
\label{smf_19}
J^{Q}_{\alpha} = \frac{1}{h} \sum\limits_{ \beta=L,R } \sum\limits_{ n = - \infty }^{\infty} \int_{-\infty}^{+\infty} d\varepsilon \, (\varepsilon_{n} - \mu) \left[  f_{\alpha}(\varepsilon_{n}) -  f_{\beta}(\varepsilon) \right] \left|S_{F,\alpha\beta}(\varepsilon_{n}, \varepsilon) \right|^2 \,,
\end{equation}
which, because of (\ref{smf_24}), is equivalent to (\ref{jqsim}).

\subsubsection{Mean power developed by the fields}
The dc power  (\ref{power}) done by the ac fields reads:
\begin{equation}
\overline{P}_l=  \frac{-i}{\tau} \int_0^{\tau}
dt \frac{d eV_l(t)}{dt}G^<_{l,l}(t,t)\,.
\end{equation} 
In terms of the representation (\ref{gfou}) it results:
\begin{eqnarray}\label{power1}
\overline{P}_{l}= \frac{\hbar\Omega_{0}\,eV_l^0}{2\pi} \sum_{n=-\infty}^{+\infty}\sum_{\alpha=L,R}
\sum_{k=\pm 1}
\int_{-\infty}^{+\infty} {d \omega} f_{\alpha}(\hbar\omega)\Gamma_{\alpha}(\omega)
\mbox{Im} \left\{ k e^{-i k \delta_l} {\cal G}_{l,l\alpha}(n, \omega )
{\cal G}_{l,l\alpha}(n+k, \omega )^* \right\}.
\end{eqnarray}
This expression does not have a counterpart in terms of the Floquet scattering matrix. This is because the evaluation of this quantity
depends on the microscopic details included explicitly in the Hamiltonian.
In fact,  notice  that the formula
(\ref{smf_24}) relates the scattering matrix only with the Green's function with the coordinates of the central system,  $l\alpha, l\beta$, that
intervene in the contacts. For the same reason, we have shown in the previous section equivalent expression within both formalisms
only for the currents through the contacts and not for the currents within $C$.  
Nevertheless the total power developed by all the fields can be also calculated within the scattering matrix formalism, see Eqs.\,(\ref{eqp}) and (\ref{smf_13}).

\subsection{Technical summary}
\begin{figure}
\includegraphics[width=0.9\columnwidth,clip]{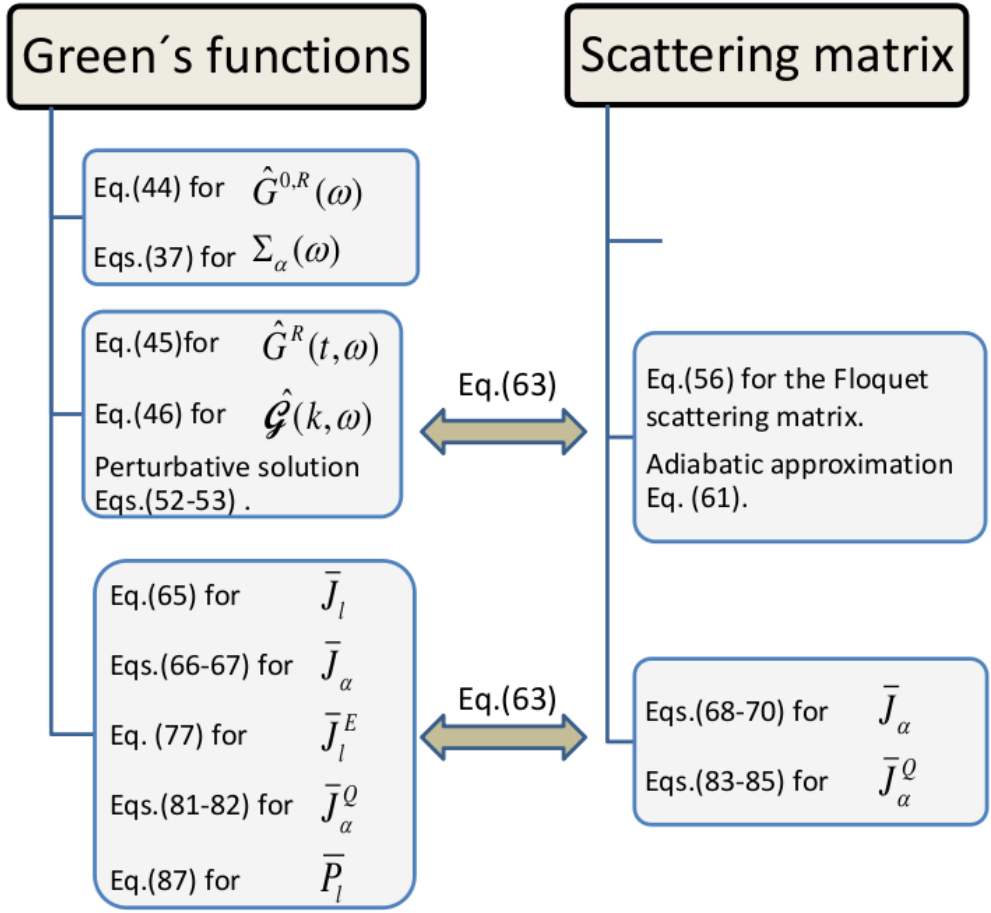}
\caption{\label{fig3} Diagram summarizing the possible steps to be followed in order to  evaluate particle, energy and heat currents as well as
the power developed by the fields by recourse to the two formalisms presented in this chapter.}
\end{figure}
To close this section we present in Figure \ref{fig3} a diagram with the summary of the procedure 
to evaluate the
different physical quantities we need to discuss the transport behavior of a
quantum pump, the alternatives and the possible approximations.

\section{Results and critical discussion}

In this section we apply the concepts and techniques
introduced in the previous section to analyze 
the conservation of the energy and the
different mechanisms of heat transport that we can identify in our quantum 
pump.
On the basis of our previous definitions we 
can show the existence of three generic effects due to a dynamical scatterer. 
At any segment of the system, it is possible to verify the conservation laws introduced in section \ref{conslaw} by
numerically solving the Dyson equation for the retarded Green's functions,
evaluating the relevant expectation values of observables following the indications
of the diagram of Figure \ref{fig3}. 
In what follows we present analytical results based on the perturbative solution of the Green's function and
the adiabatic approximation for the scattering matrix.  Without
the explicit evaluation of  the functions $\hat{G}^{0,R}(\omega)$, which depend only on the geometric
statical properties of the system, this procedure allow us to analyze
the physical properties of our system within the weak driving regime.

To make the effects  clearer we consider the case when the two reservoirs have, not only 
the same electrochemical potential $\mu_{\alpha} = \mu$, but also the same temperature
$T_{\alpha} = T~\Rightarrow$ $f_{\alpha}(\hbar\omega) = f_{0}(\hbar\omega), \forall \alpha$.

\subsection{Heating of the reservoirs by the quantum pump}

\begin{figure}
\includegraphics[width=0.9\columnwidth,clip]{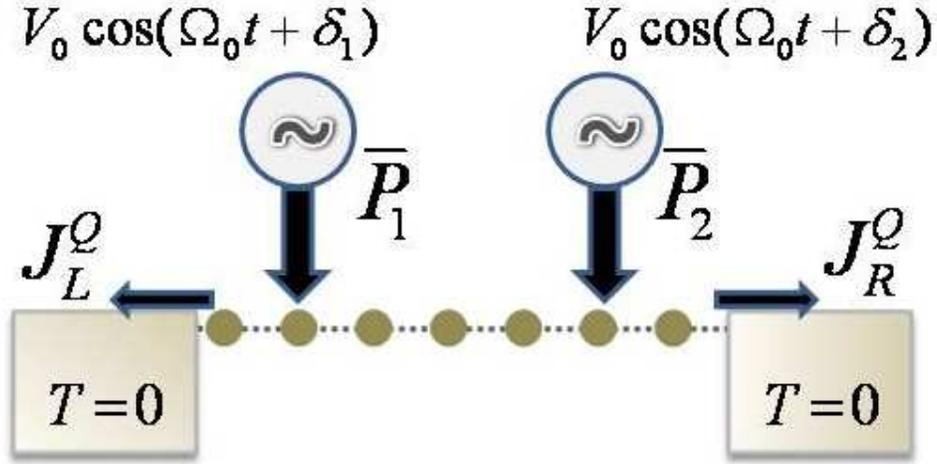}
\caption{\label{fig4} Scheme of the working regime of the quantum pump when the two reservoirs are at temperature $T=0$.
All the power developed by the external fields is dissipated in the form of heat that is absorbed by the left and right reservoirs.} 
\end{figure}

 The first effect that takes place in our quantum engine is the heating of the reservoirs 
 (see, e.g., Avron et al. 2001, Moskalets and B\"uttiker 2002, Wang and Wang 2002, 
Avron et al. 2004).
Unlike the charge current, $\overline{J}_{\alpha}$, 
the sum of heat currents in all the wires, $J^{Q}_{tot} = \sum_{\alpha} J^{Q}_{\alpha}$, 
is non zero. According to the conservation of the energy expressed in Eq.\,(\ref{conten}),
the definition of the heat current  (\ref{heatc}) and the conservation of the charge
(\ref{smf_9_1}), it is clear that the total power developed by the fields is equal
to the total heat current that enters the reservoirs:
\begin{equation}\label{eqp}
\sum_{l=1}^M \overline{P}_l = - \sum_{\alpha} J^Q_{\alpha} \,.
\end{equation}
For reservoirs at temperature $T=0$ our intuition suggests us that
the total power developed by the fields is fully transformed into heat which flows
into the reservoirs (see Figure \ref{fig4}). 
In what follows we analyze the behavior of this flow as a function of the pumping parameters
within the low driving regime. To this end, we follow an analogous procedure as
in subsection \ref{curpert}, and 
we use perturbation theory to evaluate the powers and heat flows at the 
contacts.
\subsubsection{Heat current at weak driving, $T=0$}
Following exactly the same lines as those presented in the derivation of  (\ref{jpertad}), we start from (\ref{jqsim}), we substitute
the perturbative solution of the Green's function (\ref{gfou2}) and (\ref{gper}) and expand in Taylor series the resulting expression
up to the lowest non-vanishing order in $\Omega_{0}$. The result is:
\begin{eqnarray}
J^Q_{\alpha} & \sim & \frac{\hbar \Omega_{0}^2 (eV_0)^2 }{\pi}  \sum_{j,j'=1}^M \sum_{\beta=L,R} 
\cos(\delta_j-\delta_{j'})
\Gamma_{\alpha}(\omega_{F}) \Gamma_{\beta}(\omega_{F}) \nonumber \\
& & \times 
G^{0,R}_{l\alpha,lj}(\omega_{F}) G^{0,R}_{lj,l\beta}(\omega_{F}) \left[G^{0,R}_{l\alpha,lj'}(\omega_{F}) G^{0,R}_{lj',l\beta}(\omega_{F})\right]^* .
\end{eqnarray}
The total heat flowing through the contacts reads:
\begin{eqnarray}
\sum_{\alpha=L,R} J^Q_{\alpha} & = & \frac{\hbar \Omega_{0}^2 (eV_0)^2}{\pi}  \sum_{j,j'=1}^M \sum_{\beta=L,R}
\cos(\delta_j-\delta_{j'})
\Gamma_{\alpha}(\omega_{F}) \Gamma_{\beta}(\omega_{F}) \nonumber \\
& & \times G^{0,R}_{l\alpha,lj}(\omega_{F}) G^{0,R}_{lj,l\beta}(\omega_{F}) \left[G^{0,R}_{l\alpha,lj'}(\omega_{F}) G^{0,R}_{lj',l\beta}(\omega_{F})\right]^* \nonumber \\
& = &\frac{\hbar \Omega_{0}^2 (eV_0)^2}{\pi} \sum_{j,j'=1}^M 
\cos(\delta_j-\delta_{j'}) |\rho_{lj,lj'}(\omega_{F})|^2\,, 
\end{eqnarray}
where we have used the identity between equilibrium Green's functions and the definition of the matrix presented in (\ref{propeq}).
Thus, at $T=0$ and weak driving, there is a net heat flow $\propto V_0^2 \Omega_{0}^2 $ into the reservoirs.

\subsubsection{Mean power at weak driving, $T=0$}  
We now follow a similar procedure to evaluate the mean power developed by the $j$-th force.  Substituting the perturbative solution
(\ref{gfou2}) in (\ref{power1}), and keeping terms that contribute at ${\cal O}(V_0^2)$ we get:
\begin{eqnarray}
\overline{P}_j & \sim & \frac{\hbar \Omega_{0}\, eV_0 }{\pi} \sum_{\alpha=L,R} \sum_{j'=1}^M  
\int_{-\infty}^{+\infty} d \omega f_{\alpha}(\hbar\omega) \Gamma_{\alpha}(\omega)  \nonumber \\
& & \times \mbox{Im} \left\{ e^{-i  \delta_j} [{\cal G}_{lj,l\alpha}(0,\omega)  {\cal G}^*_{lj,l\alpha}(1,\omega)  +
{\cal G}_{lj,l\alpha}(-1,\omega)  {\cal G}^*_{lj,l\alpha}(0,\omega)] \right\}.
\end{eqnarray}
Then, replacing (\ref{gper}) we derive an equation with several terms which can be collected as follows:
\begin{equation}\label{pext}
\overline{P}_j  = \sum_{j'=1}^M \left[\lambda^{(1)}_{j,j'} \cos(\delta_j-\delta_{j'}) + \lambda^{(2)}_{j,j'} \sin(\delta_j-\delta_{j'})\right]\,,
\end{equation}
with
\begin{eqnarray}
\lambda^{(1)}_{j,j'}  & = &   \frac{\hbar \Omega_{0} (eV_0)^2}{\pi} \int_{-\infty}^{+ \infty}
d \omega f_0(\hbar\omega) \mbox{Im} \left\{ \gamma_{j,j'}(\omega) \gamma_{j,j'}^{-}(\omega) \right\}\,, \nonumber \\
\lambda^{(2)}_{j,j'}  & = &  \frac{\hbar \Omega_{0} (eV_0)^2}{\pi} \int_{-\infty}^{+ \infty}
d \omega f_0(\hbar\omega) \mbox{Re} \left\{ \gamma_{j,j'}(\omega) \gamma_{j,j'}^{+}(\omega) \right\}\,,
\end{eqnarray}
being
\begin{eqnarray}
\gamma_{j,j'}(\omega) &=&  \sum_{\alpha=L,R} \left[G^{0,R}_{lj,l\alpha}(\omega)\right]^* \Gamma_{\alpha}(\omega) G^{0,R}_{lj',l\alpha}(\omega) =
-i \rho_{lj,lj'}^*(\omega)\,,\nonumber \\
\gamma_{j,j'}^{\pm}(\omega) &=& G^{0,R}_{lj,lj'}(\omega+\Omega_{0}) \pm  
G^{0,R}_{lj,lj'}(\omega-\Omega_{0})\,.
\end{eqnarray}

\subsubsection{Conservation of the energy}
The second term of (\ref{pext}) vanishes when we perform a summation over all the fields, since $\lambda_{j,j'}^{(2)}$ is symmetric
under a permutation $j \leftrightarrow j'$ while $  \sin(\delta_j-\delta_j')$ is antisymmetric under this operation.
Thus, the only term contributing to the sum over all the powers is the first one, which for low $\Omega_{0}$ results:
\begin{eqnarray}
\lambda^{(1)}_{j,j'}  & = &  \frac{\hbar \Omega_{0} (eV_0)^2}{\pi} \int_{-\infty}^{+ \infty}
d \omega \nonumber \\
& & \times
 \mbox{Re} \left\{ [f_0(\hbar\omega-\hbar\Omega_{0}) \rho_{lj,lj'}(\omega-\Omega_{0})-f_0(\hbar\omega+\hbar\Omega_{0}) \rho_{lj,lj'}(\omega+\Omega_{0})]
[G^{0,R}_{lj,lj'}(\omega)]^* \right\} \nonumber \\
 & \sim &    - \frac{2\hbar \Omega_{0}^2 (eV_0)^2}{\pi} \rho_{lj,lj'}(\omega_{F})\left[G^{0,R}_{lj,lj'}(\omega_{F})\right]^*\,.
\end{eqnarray}
Performing the sum over $j$ in (\ref{pext}) and
using $|\rho_{l,l'}(\omega)|^2= |G_{l,l'}(\omega)|^2+ |G_{l',l}(\omega)|^2-2 \mbox{Re}[G_{l,l'}(\omega)G_{l',l}(\omega)]$,
we can verify the fundamental law of the conservation of the energy (\ref{eqp}).

\subsection{Energy exchange between external forces} 
\begin{figure}
\includegraphics[width=0.9\columnwidth,clip]{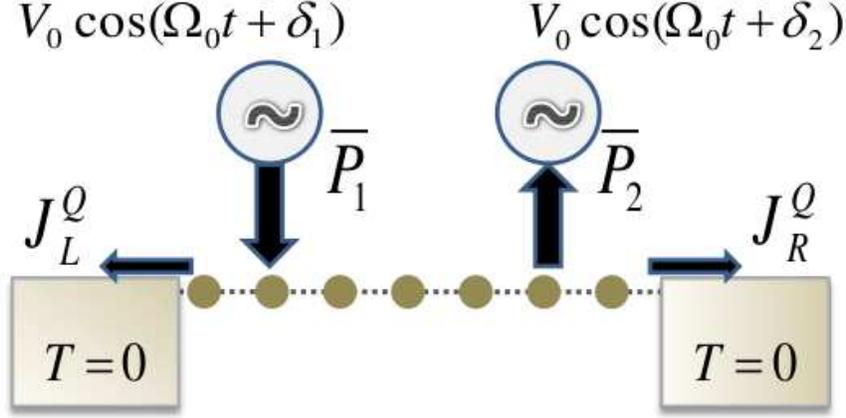}
\caption{\label{fig6} Scheme of the working regime of the quantum pump when the two reservoirs are at temperature $T=0$ and
 low driving: $V_0$ and $\Omega_0$ small. The dissipated energy flowing into the reservoirs is low, while it is possible
that part of the work done by one of the ac fields is coherently transferred to the the other one, which receives the ensuing energy.} 
\end{figure}
The evaluation of the coefficient $\lambda_{j,j'}^{(2)}$ at weak driving 
can be carried out following exactly the same steps as with $\lambda_{j,j'}^{(1)}$.
The result is 
\begin{equation}
\lambda^{(2)}_{j,j'} \sim -\frac{2\hbar \Omega_{0} (eV_0)^2}{\pi} 
\int_{-\infty}^{+\infty} d \omega \mbox{Im}\left[G^{0,R}_{lj,lj'}(\omega)G^{0,R}_{lj',lj}(\omega)\right]\,,
\end{equation}
i.e. this contribution is $\propto \Omega_{0}$, and therefore dominates the behavior
of $\overline{P}_j$ at weak driving. 
Interestingly, this contribution does not exist in
a configuration with a single ac field, while it can have different signs at different fields
in a configuration with several pumping centers. 

Therefore, we present the second general effect taking place in quantum engines: One external force can perform work directly against another external force with a negligible amount of energy being dissipated into the reservoirs.
This remarkable mechanism opens the possibility
of the coherent energy transfer between pumping centers as indicated in 
Figure \ref{fig6}.

\subsection{Directed heat transport at finite temperature}
\begin{figure}
\includegraphics[width=0.9\columnwidth,clip]{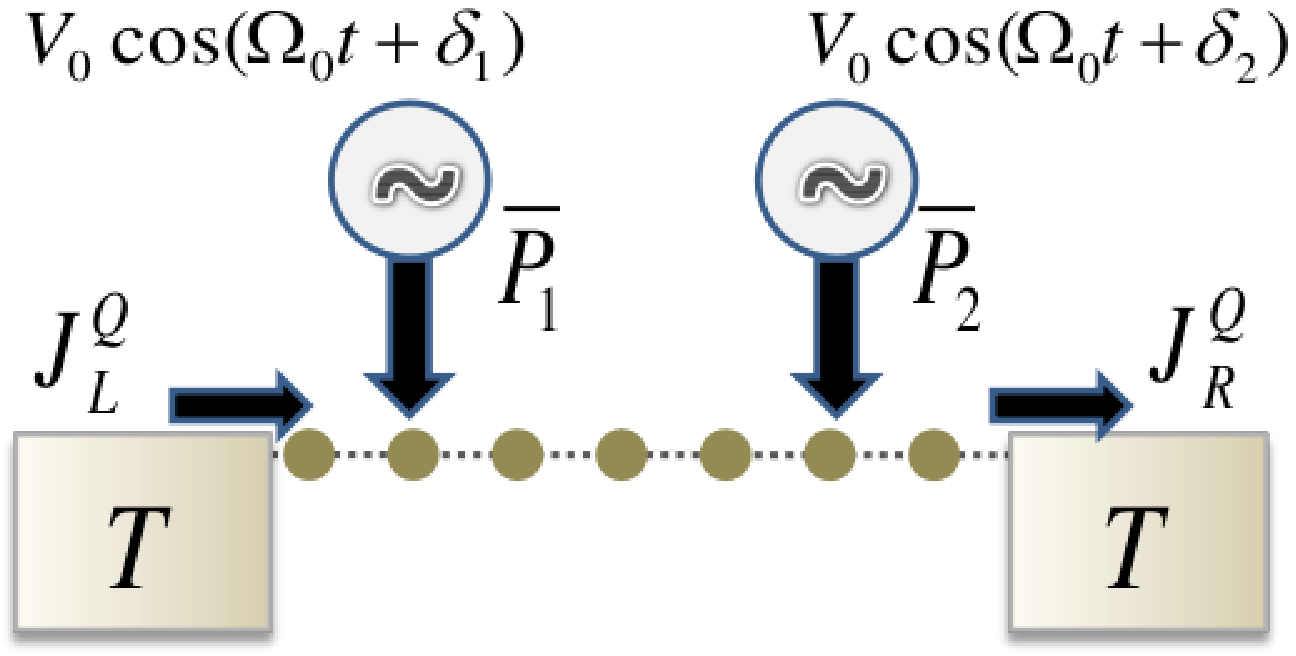}
\caption{\label{fig5} Scheme of the working regime of the quantum pump when the two reservoirs are at a finite temperature $T$.
There is a net pumping of heat from the one reservoir to the other. The quantum pump, thus works as a refrigerator.} 
\end{figure}
To show that the dynamical scatterer can induce a directed heat transfer between the reservoirs we, first, calculate the total generated heat $J^{Q}_{tot} = \sum_{\alpha} J^{Q}_{\alpha}$.
Summing up Eq.\,(\ref{smf_12}) over $\alpha$ we find (for $f_{\alpha} = f_{0}, \forall\alpha$):
\begin{equation}
\label{smf_13}
J^{Q}_{tot}  = -\,\frac{\Omega_{0} }{2\pi} \sum\limits_{ \alpha=L,R }^{} \sum\limits_{ \beta=L,R }^{} \sum\limits_{ n = - \infty }^{\infty} \int_{-\infty}^{\infty} d\varepsilon \, f_{0}(\varepsilon) \, n    \left|S_{F,\alpha\beta}(\varepsilon_{n}, \varepsilon) \right|^2 \,.
\end{equation}
The part of the total generated heat which flows into wire $\alpha$, $J^{Q}_{tot}  = \sum_{\alpha} J^{Q}_{\alpha, \, gen}$, can be defined as follows:
\begin{equation}
\label{smf_14}
J^{Q}_{\alpha, \, gen}  = -\, \frac{\Omega_{0} }{2\pi} \sum\limits_{ \beta=L,R }^{} \sum\limits_{ n = - \infty }^{\infty} \int_{-\infty}^{+\infty} d\varepsilon \, 
f_{0}(\varepsilon) \, n \left| S_{ F,\alpha\beta }( \varepsilon_{n}, \varepsilon) \right|^2 \,.
\end{equation}
The remaining part of the heat flowing into wire $\alpha$, $J^{Q}_{\alpha, \, pump} = 
J^{Q}_{\alpha} - J^{Q}_{\alpha, \, gen}$, is:
\begin{equation}
\label{smf_15}
J^{Q}_{\alpha, \, pump}  = \frac{1}{h} \int_{-\infty}^{+\infty} d\varepsilon (\varepsilon - \mu) \, f_{0}(\varepsilon) \left\{  \sum\limits_{ \beta=L,R }^{} \sum\limits_{ n = - \infty }^{\infty}  \left|S_{F,\alpha\beta}(\varepsilon_{n}, \varepsilon) \right|^2  - 1 \right\}.
\end{equation}
Using the unitarity condition for the Floquet scattering matrix, Eq.\,(\ref{smf_4}), one can easily 
show that the part of the heat current $J^{Q}_{\alpha, \, pump}$ satisfies the conservation law 
similar to the one for the charge dc current, Eq.\,(\ref{smf_9_1}): 
\begin{equation}
\label{smf_16}
\sum\limits_{\alpha = L,R } J^{Q}_{\alpha, \, pump} = 0\,.
\end{equation}
This means that $J^{Q}_{\alpha, \, pump}$ is 
transported from one reservoir to another one with the help of a dynamical scatterer. 
By analogy with the corresponding charge current we identify this portion of the total heat as a pumped heat 
(hence the lower index ``$pump$'').
This is the third general 
effect 
%mechanism 
we identified in our quantum engine: 
The dynamical scatterer induces a directed heat transport between the reservoirs (see, e.g., Humphrey et al. 2001. Segal and Nitzan 2006, Arrachea et al. 2007, Rey et al. 2007, Martinez and Hu 2007). 

If the pumped heat is, for instance, negative in the $L$ wire, $J^{Q}_{L, \, pump} < 0$, then it is necessarily positive in another wire, $J^{Q}_{R, \, pump} > 0$. 
If the absolute value of this heat is larger than the one of the
 generated component
$J^{Q}_{R, \, gen}$, then the whole heat flowing into the $R$ wire is positive, i.e. directed from the reservoir to the central system, $J^{Q}_{R} = J^{Q}_{R, \, gen} + J^{Q}_{R \, pump} > 0$. 
In this case the reservoir $R$ will be cooled while $L$ will be
  heated. 

The splitting of $J^{Q}_{\alpha}$ into $J^{Q}_{\alpha. \, gen}$ and $J^{Q}_{\alpha, \, pump}$ helped us to show that $J^{Q}_{\alpha}$ can be positive.  
Strictly speaking, such a splitting is not unique and
only the whole heat current $J^{Q}_{\alpha}$ has a direct physical meaning.
%We cannot strictly prove that such a splitting is unique.
However at slow driving, one can support such a decomposition
of $J^{Q}_{\alpha}$ into the generated  and the pumped heat by additional physical arguments as follows.

\subsubsection{Adiabatic heat currents}
The expansion (\ref{smf_17}) allows us to calculate the heat flow with an accuracy of ${\cal O}(\Omega^2)$. 
To show it explicitly we rewrite slightly Eq.\,(\ref{smf_12}) (with $f_{\alpha} = f_{0},\, \forall \alpha$). 
We assume $k_{B} T\gg \hbar\Omega_{0} $ and expand the difference of Fermi distribution functions in (\ref{smf_19})
in powers of $\Omega_{0}$, use Eq.\,(\ref{smf_17}) and find from Eq.\,(\ref{smf_19}) the heat current, $J^{Q}_{\alpha} = J^{Q,(1)}_{\alpha} + J^{Q,(2)}_{\alpha} + {\cal O}\left(\Omega_{0}^{3} \right)$, where
\begin{equation}
\label{smf_20}
J^{Q,(1)}_{\alpha} = -\,  \frac{1}{2\pi} \int_{-\infty}^{+\infty} d\varepsilon \, (\varepsilon - \mu) \left( - \frac{\partial f_{0} }{\partial \varepsilon } \right) \int_{0}^{\tau} 
\frac{dt}{\tau} \, 
\mbox{Im} \left( \hat S(t,\varepsilon) \frac{\partial \hat S^{\dagger}(t,\varepsilon) }{\partial t} \right)_{\alpha\alpha} \,, \\
\end{equation}
\begin{eqnarray}
J^{Q,(2)}_{\alpha} = -\, \frac{\hbar}{4\pi} \int_{-\infty}^{+\infty} d\varepsilon \left( - \frac{\partial f_{0} }{\partial \varepsilon } \right) \int_{0}^{\tau} \frac{dt}{\tau} \left( \frac{\partial \hat S(t,\varepsilon)}{\partial t} \frac{\partial \hat S^{\dagger}(t,\varepsilon) }{\partial t} \right)_{\alpha\alpha} \hspace{1.67cm} \nonumber \\ 
\label{smf_21}
- \frac{1}{2\pi} \int_{-\infty}^{+\infty} d\varepsilon \, (\varepsilon - \mu) \left( - \frac{\partial f_{0} }{\partial \varepsilon } \right) \int_{0}^{\tau} \frac{dt}{\tau}\, 
\mbox{Im} \left( 2\Omega_{0} \hat A(t,\varepsilon)  \frac{\partial \hat S^{\dagger}(t,\varepsilon) }{\partial t} \right)_{\alpha\alpha}  \,. 
\end{eqnarray}
Next we split the heat current into the generated heat and the pumped heat as follows, $J^{Q}_{\alpha} = J^{Q}_{\alpha, gen} + J^{Q}_{\alpha, pump}\,$, with 
\begin{equation}
\label{smf_22} 
J^{Q}_{\alpha, gen} = -\, \frac{\hbar}{4\pi} \int_{-\infty}^{+\infty} d\varepsilon \left( - \frac{\partial f_{0} }{\partial \varepsilon } \right) \int_{0}^{\tau} 
\frac{dt}{\tau} \left( \frac{\partial \hat S}{\partial t} \frac{\partial \hat S^{\dagger} }{\partial t} \right)_{\alpha\alpha},  
\end{equation}
\begin{equation}
\label{smf_23} 
J^{Q}_{\alpha, pump} = -\, \frac{1}{2\pi} \int_{-\infty}^{+\infty} d\varepsilon \, (\varepsilon - \mu) \left( - \frac{\partial f_{0} }{\partial \varepsilon } \right) \int_{0}^{\tau} \frac{dt}{\tau}\, \mbox{Im} \left( \left[ \hat S + 2\hbar\Omega_{0} \hat A \right]  \frac{\partial \hat S^{\dagger} }{\partial t} \right)_{\alpha\alpha} . 
\end{equation}
Notice that these equations also remain valid at ultralow temperatures, 
$k_{B}T \ll \hbar\Omega_{0}$, which can be verified by direct calculations taking into account the energy-independence of the matrices $\hat S$ and $\hat A$ over a scale of order $\Omega_{0}$, i.e. over the region of the thermal widening 
of the edge of the Fermi distribution function.
 
The above given splitting is justified by the following observations. (i) The quantity $J^{Q}_{\alpha, gen}$ is negative in each wire $\alpha$ as it should be for the heat generated by the scatterer and flowing into the reservoirs. (ii) At zero temperature the pumped heat vanishes identically, $J^{Q}_{\alpha, pump} = 0$, since it is impossible to take heat out of the system kept at zero temperature. 
To prove the first observation we show that the integrand in Eq.\,(\ref{smf_22}) is positive. 
To this end we use the Fourier transformation and get, $1/\tau \int_{0}^{\tau} dt (\partial\hat S/\partial t \, \partial \hat S^{\dagger}/\partial t)_{\alpha\alpha} = \Omega_{0}^2 \sum_{\beta} \sum_{n} n^2 | S_{\alpha\beta, n}|^2 > 0$. 
The second observation follows from the fact that at zero temperature it is $(\varepsilon - \mu)\,\partial f_{0}/\partial\varepsilon = 0$, hence the equation (\ref{smf_23}) vanishes.
Note that the conservation of the pumped heat current, $\sum_{\alpha} J^{Q}_{\alpha, pump} = 0$, directly follows from the conservation of charge currents, Eqs.(\ref{smf_18_1}) and (\ref{smf_18_2}), which implies $1/\tau\int_{0}^{\tau} dt\, \mbox{Im} {\rm Tr} \left[ \hat S + 2\hbar\Omega_{0} \hat A \right] \partial \hat S^{\dagger}/\partial t = 0 $.

From Eq.\,(\ref{smf_22}) it follows that the adiabatic scatterer heats the reservoirs with a rate proportional to $\hbar\Omega_{0}^2$ (Avron et al. 2001). 
In contrast, the pumped heat, Eq.\,(\ref{smf_23}), is rather proportional to $k_{B} T \Omega_{0}$. 
At sizable temperatures, $k_{B} T \gg \hbar\Omega_{0}$, the amount of pumped heat can exceed the generated heat, $|J_{\alpha, pump}|/ J_{\alpha, gen} \sim k_{B} T/(\hbar \Omega_{0}) \gg 1$.
Therefore, if in the wire $\alpha$ we have $J_{\alpha, pump} > 0$, then the reservoir $\alpha$ will be 
cooled (see Figure \ref{fig5}).
This mechanism opens the possibility of using quantum pumps as refrigerators.

\section{Summary}
In this chapter we have introduced the basic concepts to analyze at the microscopic level 
the energy transport in quantum systems driven by harmonically time-dependent fields. We
have introduced a simple microscopic model for a quantum pump, which consists in a 
finite structure connected to two macroscopic reservoirs, with ac local fields that
oscillate in time with the same frequency and a phase lag. We have analyzed the fundamental
conservation laws for the charge and the energy and we have defined the basic concepts to
study the transport behavior in these systems: charge currents, energy currents, heat currents
and powers developed by the fields. We have reviewed two complementary techniques to calculate 
the currents and the powers:
the non-equilibrium Green's function formalism for harmonically time-dependent Hamiltonians
and the scattering formalism for periodically driven mesoscopic systems. 
We have shown that the two approaches are equivalent for
the evaluation of the charge and heat currents through the contacts between the driven system 
and the reservoirs. 
We have also introduced two approximations: the adiabatic approximation to the Floquet scattering matrix and a perturbative
solution of the Dyson's equations for the Green's functions valid within the weak
driving regime. Both techniques are important
to draw conclusions on general features of the transport behavior 
%at weak driving 
without the explicit
evaluation of the Green's functions or the scattering matrix elements. Such conclusions are, thus, generic
and do not depend on the geometrical
details of the driven structure.
A summary of the technical details, including  the main equations and the alternative routes to evaluate them exactly or in an
approximate way is given in a diagram at the end of section III. 

Finally, in section IV
 we have applied the concepts and tools we have introduced in the previous sections  
in order to discuss three important mechanisms of energy transport in quantum pumps.
The first one is the fact that the total work done by all the local fields is dissipated in the form of heat that flows to the reservoirs. 
This effect is  
%completely analogous to the Joule effect and is, thus 
rather expected. 
In any case, we have exploited
our theoretical techniques at weak driving to evaluate term by term powers and heat currents and 
explicitly verify the conservation of the energy. 
To unveil a fundamental law is always a beautiful result in theoretical Physics and an important support for the power of a theoretical tool.    
In addition we have shown that other two less expected and subtle
transport mechanisms can take place: the coherent transport of energy allowing for regimes where
some of the forces make work, while other receive work. This interesting mechanism could be
exploited, for instance, to couple two quantum pumps in a combined engine. The final 
remarkable mechanism is the pumping of heat at finite temperature and weak driving, allowing
for the operation of the quantum pump as a refrigerator which extracts heat from a reservoir
and injects heat in the other one.

\section{Future perspective}
The different operational regimes that we have identified in the quantum pumps have several 
important outcomes. On the theoretical side there are several lines to further analyze.
A first issue to explore is the role of the geometrical details of the structure, 
in order to identify the optimal architecture to enhance each mechanism and improve the
efficiency of the quantum engine. Another important ingredient is the investigation of
the role of many-body interactions. In particular, the electron-electron and the 
electron-phonon interactions. On the   
experimental side it would be very interesting the design of an
 experimental setup to implement these effects. In this sense, it is very promising 
  that quantum refrigeration has been already experimentally
explored in mesoscopic structures with superconducting
elements under ac driving [Giazzoto 2006]. 

\section{Acknowledgments}
We thank Luis Martin-Moreno for useful discussions and C. Marcus for Figure 1.
 LA acknowledges support from CONICET and UBACyT, Argentina.

\section{Bibliography}

%\begin{thebibliography}

%\bibitem{ar2002}
Arrachea, L. 2002.
Current oscillations in a metallic ring threaded by a time-dependent magnetic flux.
Physical Review B 66: 045315 (11).

%\bibitem{ar2005}
Arrachea, L. 2005.  A Green-function approach to transport phenomena in quantum pumps,
Physical Review B 72: 125349 (11). 

%\bibitem{ArracheaMoskalets2006}
Arrachea, L. and Moskalets, M. 2006. Relation between scattering matrix and Keldysh formalisms for quantum transport driven by time-periodic fields. Physical Review B 74: 245322 (13).

%\bibitem{AMM2007}
Arrachea, L., Moskalets, M., and Martin-Moreno, L. 2007. Heat production and energy balance in nanoscale engines driven by time-dependent fields. Physical Review B 75: 245420 (5).

%\bibitem{AEGS2001}
Avron, J. E.,  Elgart, A., Graf, G. M., and Sadun, L. 2001. Optimal Quantum Pumps. Physical Review Letter 87: 236601 (4).

%\bibitem{AEGS2004}
Avron, J. E., Elgart, A., Graf, G. M., and Sadun, L. 2004. Transport and Dissipation in Quantum Pumps. Journal of Statistical Physics 116: 425 - 73.

%\bibitem{Brouwer1998}
Brouwer, P. W. 1998. Scattering approach to parametric pumping. Physical Review B 58: R10135 - 8.

%\bibitem{Buttiker1990}
B\"uttiker, M. 1990. Scattering theory of thermal and excess noise in open conductors. Physical Review Letter 65: 2901 - 4.

%\bibitem{Buttiker1992}
B\"uttiker, M. 1992. Scattering theory of current and intensity noise correlations in conductors and wave guides. Physical Review B 46: 12485 - 507.

%\bibitem{Buttiker1993}
B\"uttiker, M. 1993. Capacitance, admittance, and rectification properties of small conductors. Journal of Physics: Condensed Matter 5: 9361 - 78.

%\bibitem{BTP1994}
B\"uttiker, M., Thomas, H., and Pr\^{e}tre, A. 1994. Current partition in multiprobe conductors in the presence of slowly oscillating external potentials. Zeitschrift f\"ur Physik B Condensed Matter 94: 133 - 7.

%\bibitem{} 
Caroli, C, Combescot, R., Nozieres, P. and Saint-James, D. 1971. Direct calculation of the tunneling current. Journal of Physics C: Solid State Physics 4: 916 - 929.

%\bibitem{FisherLee1981}
Fisher, D. S. and Lee, P. A. 1981. Relation between conductivity and transmission matrix. Physical Review B 23: 6851 - 4.  

Giazotto, F., Heikkila, T.T., Luukanen, A., Savin, A.M., Pekola, J.P. 2006. Opportunities for mesoscopics in thermometry and refrigeration: Physics and applications. Rev. Mod. Phys. 78: 217 - 274.

%\bibitem{haug}
Haug, H. and Jauho, A. P. 1996. Quantum Kinetics in Transport and Optics in Semiconductors: Springer Solid-State Sciences 123.

%\bibitem{HLN2001}
Humphrey, T. E., Linke, H., and Newbury, R. 2001. Pumping heat with quantum ratchets. Physica E 11: 281 - 6.

%\bibitem{kada}
Kadanoff, L. P, Baym, G. 1962. Quantum statistical mechanics: Benjamin/Cummings Publishing Group USA.

%\bibitem{Keldysh}
Keldysh, L. V. 1964. Diagram technique for nonequilibrium processes. Zh. Eksp. Teor. Fiz. 47: 1515 - 27.

Kohler, S., Lehmann, J., H\"{a}nggi, P. 2005. Driven quantum transport on the nanoscale. Physics Reports 406: 379 - 443. 

%\bibitem{jau}
Jauho, A. P., Wingreen, N, and Meir, Y. 1994. Time-dependent transport in interacting and noninteracting resonant-tunneling systems.
Physical Review B 50: 5528 - 44.

%\bibitem{Landauer1957}
Landauer, R. 1957. Spatial Variation of Currents and Fields Due to Localized Scatterers in Metallic Conduction. IBM Journal of Research and Development 1: 223 - 31.

%\bibitem{Landauer1970}
Landauer, R. 1970. Electrical resistance of disordered one-dimensional lattices. Philosophical Magazine 21: 863 - 7.

%\bibitem{Landauer1975}
Landauer, R. 1975. Residual Resistivity Dipoles. 
Zeitschrift f\"ur Physik B Condensed Matter 21: 247 - 54.
%Z. Phys. B 21: 247 - 54.

%\bibitem{Mahan}
Mahan, G. D. 1990. Many Particle Physics. New York: Plenum.

%\bibitem{MartinezHu2007}
Martinez, D. F. and Hu, B. 2007. Operating molecular transistors as heat pumps. arXiv:0709.4660v1.

%\bibitem{MoskaletsButtiker2002}
Moskalets, M. and B\"uttiker, M. 2002a. Dissipation and noise in adiabatic quantum pumps. Physical Review B 66: 035306 (9).

%\bibitem{MoskaletsButtiker2002a}
Moskalets, M. and B\"uttiker, M. 2002. Floquet scattering theory of quantum pumps. Physical Review B 66: 205320 (10).

%\bibitem{MoskaletsButtiker2004}
Moskalets, M. and B\"uttiker, M. 2004. Adiabatic quantum pump in the presence of external ac voltages. Physical Review B 69: 205316 (12).

%\bibitem{MoskaletsButtiker2005}
Moskalets, M. and B\"uttiker, M. 2005. Magnetic field symmetry of pump currents of adiabatically driven mesoscopic structures. Physical Review B 72: 035324 (11). 

%\bibitem{past}
Pastawski, H. 1992. Classical and quantum transport from generalized Landauer-B\"uttiker equations. II. Time-dependent resonant tunneling, Physical Review B 46: 4053 - 70.

%\bibitem{PlateroAguado2004}
Platero, G. and Aguado, R. 2004. Photon-assisted transport in semiconductor nanostructures. Physics Reports 395: 1 - 157.

%\bibitem{RSKH2007}
Rey, M., Strass, M., Kohler, S., H\"anggi, P., and Sols, F. 2007. Nonadiabatic electron heat pump. Physical Review B 76: 085337 (4).

%\bibitem{schwin}
Schwinger, J. 1961. Brownian Motion of a Quantum Oscillator. Journal of Mathematical Physics 2: 407 - 432.

%\bibitem{SegalNitzan2006}
Segal, D., Nitzan, A. 2006. Molecular heat pump. Physical Review E 73: 026109 (9).

%\bibitem{WW2002}
Wang, B., Wang, J. 2002. Heat current in a parametric quantum pump. Physical Review B 66: 125310 (4).

%\end{thebibliography}

%\listoffigures

\end{document}